\documentclass[acmsmall]{acmart}
\AtBeginDocument{%
  }

\setcopyright{acmlicensed}
\copyrightyear{2018}
\acmYear{2025}
\acmDOI{XXXXXXX.XXXXXXX}

\usepackage{enumitem}
\usepackage{amsmath}
\usepackage{booktabs}
\usepackage{multirow}
\usepackage{adjustbox}
\usepackage{graphicx}
\usepackage{caption}
\usepackage{tcolorbox}
\usepackage{threeparttable}

\newcommand{\consolaslike}[1]{\texttt{\textcolor[HTML]{078080}{\ttfamily#1}}}
\begin{document}

\settopmatter{printacmref=false}

\title{Beyond Basic Specifications? A Systematic Study of Logical Constructs in LLM-based Specification Generation}



\author{Zehan Chen}
\affiliation{%
  \institution{School of Computer Science and Technology, Xidian University}
  \city{Xi'an, Shannxi}
  \country{China}}
\email{zehanchen@stu.xidian.edu.cn}

\author{Long Zhang}
\affiliation{%
  \institution{National Key Laboratory of Science and Technology on Information System Security, AMS}
  \city{Beijing}
  \country{China}}
\email{zhanglong10@nudt.edu.cn}

\author{Zhiwei Zhang}
\affiliation{%
  \institution{School of Computer Science and Technology, Xidian University}
  \city{Xi'an, Shannxi}
  \country{China}}
\email{zwzhang@xidian.edu.cn}

\author{JingJing Zhang}
\affiliation{%
  \institution{Unit 32092 of PLA}
  \city{Beijing}
  \country{China}}
\email{lg02103@163.com}

\author{Ruoyu Zhou}
\affiliation{%
  \institution{School of Computer Science and Technology, Xidian University}
  \city{Xi'an, Shannxi}
  \country{China}}
\email{23031212412@stu.xidian.edu.cn}

\author{Yulong Shen}
\affiliation{%
  \institution{School of Computer Science and Technology, Xidian University}
  \city{Xi'an, Shannxi}
  \country{China}}
\email{ylshen@mail.xidian.edu.cn}

\author{JianFeng Ma}
\affiliation{%
  \institution{School of Cyber Engineering, Xidian University}
  \city{Xi'an, Shannxi}
  \country{China}}
\email{jfma@mail.xidian.edu.cn}

\author{Lin Yang}
\authornote{Lin Yang is the corresponding author.}
\affiliation{%
  \institution{National Key Laboratory of Science and Technology on Information System Security, Systems Engineering Institute, AMS}
  \city{Beijing}
  \country{China}}
\email{yanglin61s@126.com}

\renewcommand{\shortauthors}{Chen et al.}

\begin{abstract}
Formal specifications play a pivotal role in accurately characterizing program behaviors and ensuring software correctness. In recent years, leveraging large language models (LLMs) for the automatic generation of program specifications has emerged as a promising avenue for enhancing verification efficiency. However, existing research has been predominantly confined to generating specifications based on basic syntactic constructs, falling short of meeting the demands for high-level abstraction in complex program verification. Consequently, we propose incorporating logical constructs into existing LLM-based specification generation framework. Nevertheless, there remains a lack of systematic investigation into whether LLMs can effectively generate such complex constructs. To this end, we conduct an empirical study aimed at exploring the impact of various types of syntactic constructs on specification generation framework. Specifically, we define four syntactic configurations with varying levels of abstraction and perform extensive evaluations on mainstream program verification datasets, employing a diverse set of representative LLMs. Experimental results first confirm that LLMs are capable of generating valid logical constructs. Further analysis reveals that the synergistic use of logical constructs and basic syntactic constructs leads to improvements in both verification capability and robustness, without significantly increasing verification overhead. Additionally, we uncover the distinct advantages of two refinement paradigms. To the best of our knowledge, this is the first systematic work exploring the feasibility of utilizing LLMs for generating high-level logical constructs, providing an empirical basis and guidance for the future construction of automated program verification framework with enhanced abstraction capabilities.
\end{abstract}

\ccsdesc[500]{Software and its engineering~Software verification}

\keywords{Specification Generation, Large Language Models, Logical Constructs}


\maketitle

\section{Introduction}
In program verification, formal specifications are used to precisely characterize program behaviors and the properties that programs are required to satisfy, thereby mitigating the ambiguity that often arises when software requirements are described in natural language~\cite{Kamsties-01,Shah-01}. However, traditional specification generation methods typically focus on specific types of specifications, which limits their generality and prevents them from generating multiple kinds of specifications simultaneously~\cite{Ryan-01,Si-01,Yao-01}. Recent studies~\cite{Kamath-01,Ma-01,Wen-01} have sought to address this limitation by introducing large language models (LLMs). Trained on large-scale code corpora, these models can automatically generate candidate specifications for given programs, demonstrating promising potential for specification generation.

\textbf{Research Gap}. Although these approaches exhibit certain advantages, they are currently limited to generating specifications that contain only basic syntactic constructs. This limitation makes the resulting specifications incapable of expressing complex mathematical concepts and program structures, thereby depriving verification tools of necessary abstraction information and hindering the verification of complex programs. In fact, most specification languages~~\cite{Burdy-01, Leino-01, Patrick-01} provide advanced syntactic features, such as various logical constructs—including logical functions, lemmas, and axioms—to support modular reasoning and proof guidance. However, whether LLMs can be leveraged to generate more abstract logical constructs, and how different types of such constructs, when generated by LLMs, affect the verification capability, stability, and efficiency of the overall specification generation framework, have not yet been systematically studied. This research gap motivates the present work.

\textbf{Goal}. Therefore, our goal is to bridge this gap by investigating the performance of incorporating logical constructs into existing LLM-based specification generation framework. We aim to address the following research questions (RQs):
\begin{itemize}[leftmargin=*, labelsep=0.5em]
    \item \textbf{RQ1. Can current LLM-based specification generation framework effectively generate and utilize logical constructs to support program verification?} Compared to basic syntactic constructs, logical constructs typically impose stricter syntactic and semantic constraints, thereby placing higher demands on the structured output capabilities of generative models. Consequently, before discussing the specific performance of different types of syntactic constructs, it is necessary to first assess whether LLMs can reliably adhere to a predefined subset of the specification language syntax under prompt constraints, and generate logical constructs that are both syntactically valid and acceptable to verification tools. This RQ aims to evaluate the feasibility of such generation capabilities, laying the foundation for subsequent analyses of performance and efficiency.

    \item \textbf{RQ2. How do different types of syntactic constructs differ in their performance in program verification?} Basic syntactic constructs are generally easier to generate, but their expressive power is limited. In contrast, logical constructs can raise the level of abstraction and enhance expressiveness; however, their practical impact on program verification remains unclear. Moreover, there is currently a lack of systematic empirical studies that compare the overall performance of these constructs in terms of verification success rate, verification stability, and verification efficiency. This RQ addresses this gap by conducting a comparative analysis of different syntactic constructs, aiming to reveal the trade-offs between specification generation capability and efficiency.

    \item \textbf{RQ3. How do different types of syntactic constructs behave under different specification refinement paradigms?} Specification refinement is a critical factor affecting both final verification outcomes and verification efficiency; however, existing studies have not systematically analyzed the sensitivity of different types of syntactic constructs to refinement strategies. Accordingly, this RQ aims to investigate the interaction between syntactic construct types and specification refinement paradigms, thereby elucidating their combined impact on verification success rate, number of refinement iterations, and overall efficiency.
\end{itemize}

\textbf{Methodology}. To address the above RQs, we propose a comprehensive evaluation framework to investigate the behaviors of different syntactic constructs in specification languages within current specification generation framework. Specifically, we first design four syntactic construct configurations using different prompts and examples: a configuration based on basic syntactic constructs, a configuration based on verifiable logical constructs, a configuration based on axiom constructs, and an unconstrained configuration. Subsequently, under two refinement paradigms, we evaluate these four configurations using six LLMs with varying capability levels, including both commercial and open-source models from leading companies. Finally, based on the experimental data, we conduct a systematic analysis of the RQs using multidimensional evaluation metrics.

\textbf{Findings}. Based on extensive empirical studies, we summarize the following key findings: (1) The vast majority of mid-to-high-capability LLMs are capable of generating logical specifications adhering to syntactic configurations. Furthermore, we observe that high-capability models may proactively avoid complex logical expressions to ensure correctness. (2) The relationship between logical constructs and basic syntactic constructs is not hierarchical replacement but rather exhibits strong complementarity. The synergistic use of diverse syntactic constructs leverages the strengths of both, significantly enhancing the overall verification capability of the framework. (3) Introducing logical constructs increases the uncertainty of the verification process. However, combining basic syntactic constructs with logical constructs effectively mitigates this instability in high-capability models. (4) Although logical constructs introduce additional complexity, their synergy with basic syntax does not significantly increase the exploration cost of verification tools or runtime, thereby demonstrating the feasibility of introducing high-level abstractions. (5) There is no absolutely dominant refinement strategy; the deletion paradigm holds an advantage in verification efficiency, while the modification paradigm offers superior correction capability and specification expressiveness.

In summary, this paper makes the following contributions:
\begin{itemize}[leftmargin=*, labelsep=0.5em]
    \item To the best of our knowledge, this work presents the first systematic study investigating the feasibility of using LLMs to generate advanced logical constructs, thereby providing new insights into LLM-based logical abstraction for specification generation.
    \item Building upon existing specification generation framework, we conduct a comprehensive evaluation of four syntactic construct configurations under different specification refinement paradigms, and identify important directions for future research.
    \item The code and data used in this study have been publicly released.
\end{itemize}

\section{Background and Motivation}
\subsection{Specification Language}
\begin{figure}[htbp]
  \centering
  
  \begin{minipage}{0.5\linewidth}
    \centering
    \includegraphics[width=\linewidth]{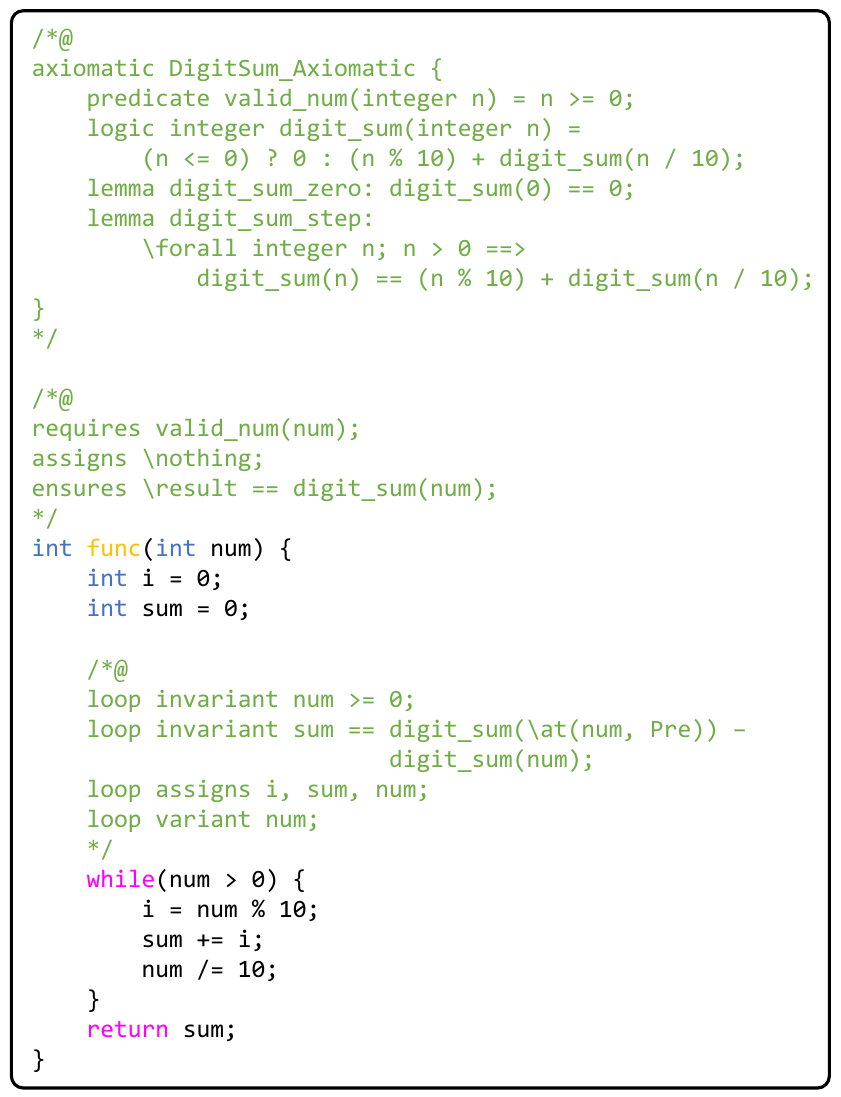}
    \caption*{(a) Verifiable Logical Constructs.}
  \end{minipage}%
  \begin{minipage}{0.5\linewidth}
    \centering
    \includegraphics[width=\linewidth]{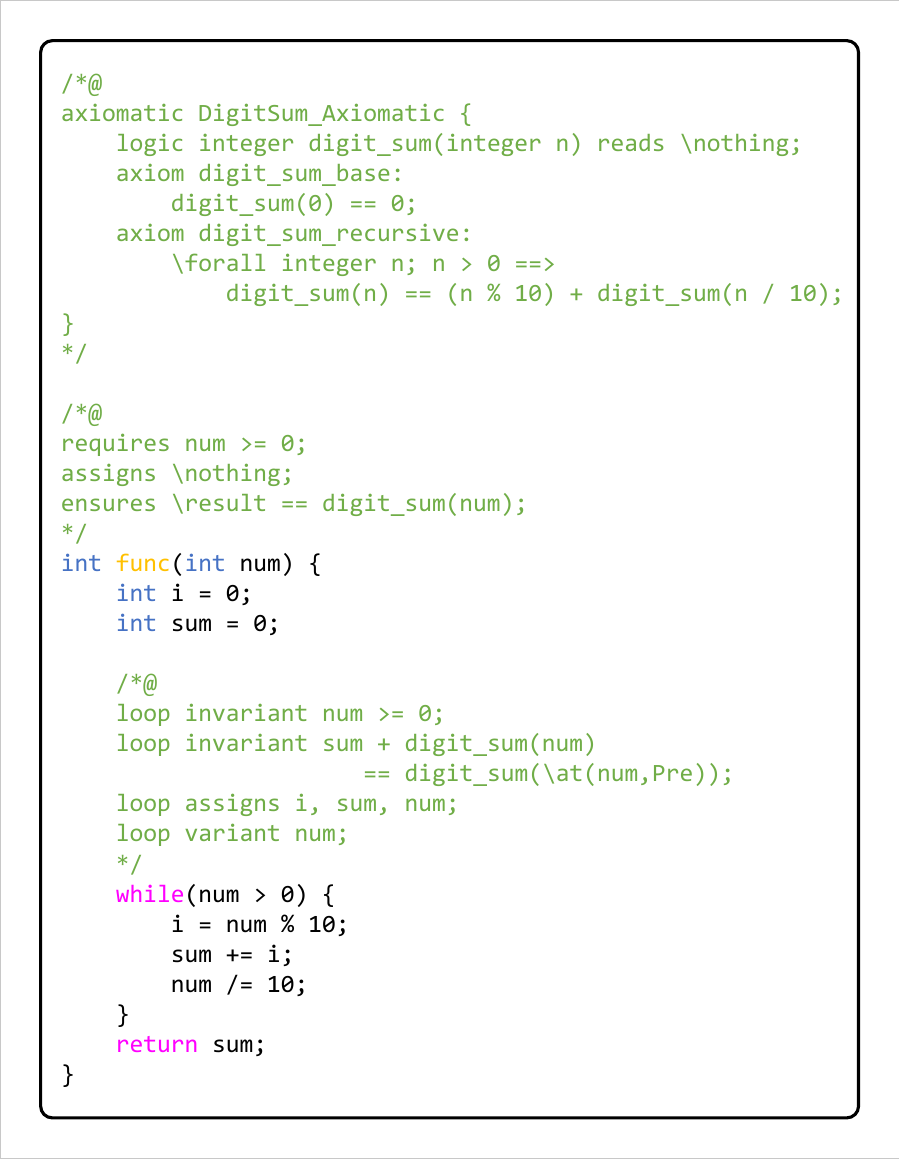}
    \caption*{(b) Axiom Constructs.}
  \end{minipage}%
  
  \caption{An Example of Specifications Written with Verifiable Logical Constructs and Axiom Constructs, Respectively.}
  \label{fig: Example}
\end{figure}

Formal specification languages are grounded in mathematical and logical formalisms to precisely describe the intended behaviors of programs and to rigorously define the properties that programs must satisfy. Different programming languages are associated with different formal specification languages~\cite{Burdy-01, Leino-01, Patrick-01}. In this study, we focus on the specification language for C programs—the ANSI/ISO C Specification Language (ACSL)~\cite{Patrick-01}—which is embedded in source code in the form of annotations. Fig.~\ref{fig: Example} illustrates an example of a program annotated with ACSL specifications. In ACSL, \consolaslike{requires} specifies the properties expected to hold for function inputs, \consolaslike{ensures} states the properties that must hold after the function has completed execution, and \consolaslike{loop invariant} specifies the properties that must be preserved before and after each loop iteration. The clauses \consolaslike{assigns} and \consolaslike{loop assigns} specify the memory locations that may be modified during the execution of a function and a loop, respectively. To prove loop termination, \consolaslike{loop variant} can be used to specify a termination measure. In addition to these basic syntactic constructs, ACSL provides more advanced logical constructs. The constructs \consolaslike{predicate} and \consolaslike{logic} are used to define pure functions that can be used only within specifications; both can take different labels and values as parameters. The key difference is that \consolaslike{predicate} can return only Boolean values, whereas \consolaslike{logic} can return values of any specified type. The construct \consolaslike{lemma} is used to state general properties of predicates and logic functions, which must be independently proved by the verification tool. Once proved, these properties can be safely used to simplify reasoning in other, more complex proofs without requiring re-proving. The construct \consolaslike{axiom} is used to state properties that are declared to hold in all cases. By abstracting complex properties, axioms can make the proof process more efficient; however, since any property expressed as an axiom is assumed to be true, they must be used with great caution, as improper use can invalidate the entire verification process.

\subsection{LLM-Based Specification Generation Framework}
In recent years, related research has widely adopted a "guess–verify" framework for program invariant generation~\cite{Hojjat-01, Nguyen-02, Ryan-01, Xu-01, Yao-01, Yu-01}. In this framework, specifications are generated in the guessing phase and then verified using formal tools in the verification phase. Building upon prior work such as \cite{Kamath-01} and \cite{Wu-01}, this paper formally summarizes and describes an LLM-based specification generation framework consisting of three phase: guess, verify, and refine. Specifically, in the guessing phase, the program $\mathcal{P}$ and the generation prompt $\mathcal{M}_{\text{generate}}$ are jointly provided to an LLM oracle $\mathcal{O}_{\text{propose}}$. The oracle $\mathcal{O}_{\text{propose}}$ generates a set of candidate specifications $\mathcal{S}$ according to the prompt, i.e., $\mathcal{S} = \mathcal{O}_{\text{propose}}(\mathcal{P}, \mathcal{M}_{\text{generate}})$. In the verification phase, the program $\mathcal{P}$, the candidate specification set $\mathcal{S}$, and the property $p$ to be verified are jointly fed into a verification tool $\mathcal{V}$. Different actions are taken based on the verification result. If verification succeeds (i.e., $\mathcal{V}(\mathcal{P}, \mathcal{S}, p) = \text{True}$), the specification set $\mathcal{S}$ is directly returned as the final specifications. Otherwise (i.e., $\mathcal{V}(\mathcal{P}, \mathcal{S}, p) = \text{False}$), the framework proceeds to the refinement phase. In the refinement phase, the framework refines the specifications based on the feedback information $\mathcal{I}$ produced by the verification tool $\mathcal{V}$, and iteratively repeats this process until the program is successfully verified or a predefined maximum number of iterations is reached. Currently, there are two main refinement paradigms: (1) \textbf{Deletion paradigm}~\cite{Wen-01, Chen-01}, which directly removes erroneous specifications. The new specification set is given by $\mathcal{S}' = \mathcal{S} \setminus \mathcal{S}_{\text{error}}$, where $\mathcal{S}_{\text{error}} = \mathcal{SA}(\mathcal{I})$ denotes the set of erroneous specifications obtained by performing static analysis $\mathcal{SA}$ on the feedback information $\mathcal{I}$. (2) \textbf{Modification paradigm}~\cite{Ma-01}, which regenerates specifications by jointly feeding the program $\mathcal{P}$, the specification set $\mathcal{S}$, the feedback information $\mathcal{I}$, and a repair prompt $\mathcal{M}_{\text{repair}}$ into an LLM oracle $\mathcal{O}_{\text{repair}}$. The new specification set is thus given by $\mathcal{S}' = \mathcal{O}_{\text{repair}}(\mathcal{P}, \mathcal{S}, \mathcal{I}, \mathcal{M}_{\text{repair}})$.

\subsection{Motivating Example}
The function \textit{func} in Fig.~\ref{fig: Example} computes the sum of the decimal digits of the input integer \textit{num}. We observe that it is difficult to write specifications that accurately characterize the behavior of this function using only basic syntactic constructs. This difficulty arises because the keywords provided by basic constructs are primarily suited to describing relatively direct and simple relationships between inputs and outputs, and are inadequate for formalizing more complex mathematical concepts such as decimal decomposition and digit-wise accumulation. To address this limitation, one can leverage the logical constructs provided by ACSL. As shown in Fig.~\ref{fig: Example}-(A), a logic function \textit{digit\_sum} can be written to recursively model the implementation of \textit{func}, while the lemmas \textit{digit\_sum\_zero} and \textit{digit\_sum\_step} describe the general properties of \textit{digit\_sum} when the input parameter $n$ equals 0 and when it is greater than 0, respectively. Fig.~\ref{fig: Example}-(B) presents a more concise formulation. Specifically, instead of providing a concrete implementation of \textit{digit\_sum}, its behavior is specified using the axioms \textit{digit\_sum\_base} and \textit{digit\_sum\_recursive}, which do not require verification. By using these logical constructs, it becomes significantly simpler and faster to write specifications for complex functions—something that is difficult to achieve using basic syntactic constructs alone. This observation motivates our interest in introducing logical constructs into existing LLM-based specification generation framework.

\section{Evaluation Framework}
\begin{figure*}[tbph]
    \centering
    \includegraphics[width=\textwidth]{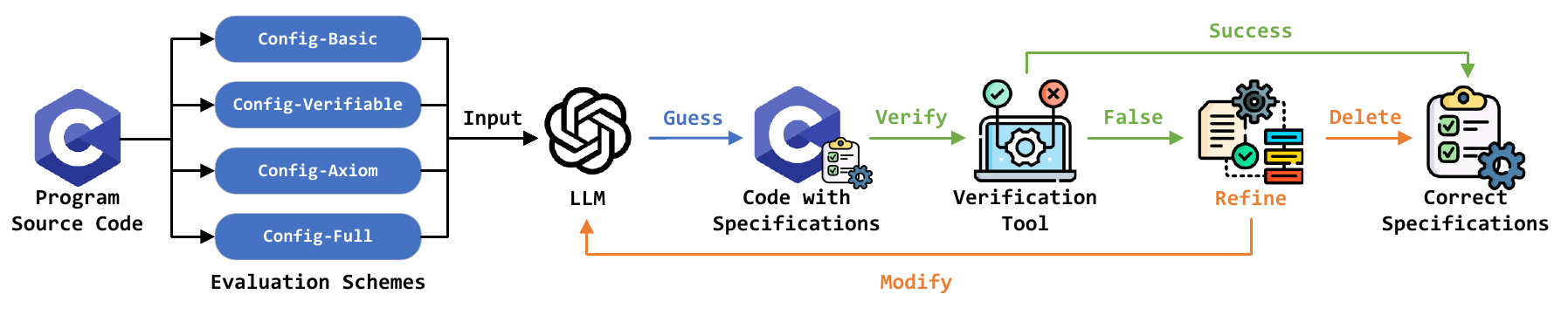}
    \caption{Evaluation Framework.}
    \label{pic: Evaluation Framework}
    \Description{}
\end{figure*}

Fig~\ref{pic: Evaluation Framework} presents an overview of our syntactic construct evaluation framework, whose execution flow follows existing LLM-based specification generation framework. In the following, we describe the concrete implementation details of this framework.

\subsection{Syntactic Construct Configurations}
We define a configuration set $\mathcal{C} = \{$\text{Config-Basic (CB)}, \text{Config-Verifiable (CV)}, \text{Config-Axiom (CA)}, \text{Config-Full (CF)}$\}$, where each element represents a distinct syntactic construct configuration. For any configuration $c \in \mathcal{C}$, there is a corresponding subset of syntactic constructs $\mathcal{P}_c$ that are permitted under that configuration. In addition, for CV and CA, we further define a mandatory construct subset $\mathcal{R}_c$. The concrete definitions of the four configurations are as follows:
\begin{itemize}[leftmargin=*, labelsep=0.5em]
    \item \textbf{CB}: This configuration serves as the baseline. It prompts the oracle $\mathcal{O}_{\text{propose}}$ to use only the basic syntactic constructs required to describe program behaviors, thereby simulating existing specification generation approaches such as AutoSpec~\cite{Wen-01} and SpecGen~\cite{Ma-01}. The permitted construct subset is $\mathcal{P}_{\text{Basic}} = \{$\texttt{requires}, \texttt{ensures}, \texttt{assigns}, \texttt{loop invariant}, \texttt{loop variant}, \texttt{loop assigns}, \texttt{behavior}$\}$.
    \item \textbf{CV}: This configuration extends the baseline by introducing strictly verifiable logical constructs. It encourages the oracle $\mathcal{O}_{\text{propose}}$ to abstract complex functional semantics in the program into predicates and logic functions, and to generate additional lemmas to assist verification. Unlike CA, these constructs must be automatically discharged by the verification tool. The mandatory construct subset is $\mathcal{R}_{\text{Verifiable}} = \{\texttt{predicate}, \texttt{logic}, \texttt{lemma}\}$, and the corresponding permitted construct subset is $\mathcal{P}_{\text{Verifiable}} = \mathcal{P}_{\text{Basic}} \cup \mathcal{R}_{\text{Verifiable}}$.
    \item \textbf{CA}: This configuration introduces axiom-related syntactic constructs, allowing the oracle $\mathcal{O}_{\text{propose}}$ to define logical rules that can be used without being automatically proved by the verification tool. The mandatory construct subset is $\mathcal{R}_{\text{Axiom}} = \{\texttt{axiom}\}$, and the corresponding permitted construct subset is $\mathcal{P}_{\text{Axiom}} = \mathcal{P}_{\text{Basic}} \cup \{\texttt{predicate}, \texttt{logic}\} \cup \mathcal{R}_{\text{Axiom}}$.
    \item \textbf{CF}: This configuration characterizes the capability of the oracle $\mathcal{O}_{\text{propose}}$ in the absence of any syntactic restrictions. In this setting, $\mathcal{O}_{\text{propose}}$ is allowed to freely employ arbitrary syntactic constructs to achieve the verification goal. By combining strictly verifiable logical constructs with axiomatic assumptions, this configuration simulates the scenario in which experienced verification engineers leverage all available mechanisms to complete code verification. The permitted construct subset is $\mathcal{P}_{\text{Full}} = \mathcal{P}_{\text{Verifiable}} \cup \mathcal{P}_{\text{Axiom}}$.
\end{itemize}

For any configuration $c \in \mathcal{C}$, we construct a corresponding prompt $\mathcal{M}_{\text{generate}}^{c}$ in the guessing phase. This prompt imposes configuration-specific syntactic constraints on the oracle $\mathcal{O}_{\text{propose}}$, ensuring that the syntactic constructs used in the generated candidate specifications $S$ belong to the subset associated with that configuration. Formally, we have $\mathcal{S} = \mathcal{O}_{\text{propose}}(\mathcal{P}, \mathcal{M}_{\text{generate}}^{c})$ \text{s.t.} $\text{Constr}(\mathcal{S}) \subseteq \mathcal{P}_c$, where $\text{Constr}(\mathcal{S})$ denotes the set of syntactic constructs used in specifications $\mathcal{S}$. In addition, for CV and CA, an extra constraint must be satisfied: $\text{Constr}(\mathcal{S}) \cap \mathcal{R}_c \neq \emptyset$.

\subsection{Dataset}
In this study, we use the \textit{frama-c-problems}\cite{Manav-01} dataset as the primary benchmark for experimental evaluation. This dataset is one of the most widely adopted standard test suites in research on ACSL-based specification generation and has been used as the main evaluation benchmark in prior work such as AutoSpec\cite{Wen-01} and SLD-Spec\cite{Chen-01}. As a result, it ensures both the comparability and reproducibility of our experimental results. The dataset comprises a collection of representative C program examples, which are organized into eight categories and cover a wide range of program characteristics—from basic arithmetic computations, pointer manipulations, and array operations to more complex algorithms such as sorting. Furthermore, to enhance experimental rigor, we extend and clean the dataset based on the version used in AutoSpec. These modifications include enriching the calling contexts of target functions, thereby forcing the generated specifications to exhibit a higher degree of completeness, and removing duplicate or semantically meaningless examples. The resulting refined dataset contains a total of 49 program instances.

\subsection{Studied LLMs}
To comprehensively evaluate the generalization ability and overall performance of the four syntactic construct configurations across different LLMs, we conduct experiments using a set of representative state-of-the-art (SOTA) models. These models span both open-source and closed-source categories, as well as general-purpose foundation models and models specifically optimized for code-related tasks. On the closed-source side, we evaluate OpenAI’s GPT‑3.5‑Turbo~\cite{gpt-3.5-turbo}, GPT‑4o~\cite{gpt-4o}, and GPT‑5~\cite{gpt-5}, as well as Google’s Gemini‑2.5‑Pro~\cite{gemini-2.5-pro}. On the open-source side, we select Qwen2.5‑Coder‑32B~\cite{Hui-01} and Llama‑3.1‑70B~\cite{meta-llama-3-1}, which have demonstrated strong performance on multiple code generation benchmarks.

\subsection{Evaluation Metrics}
We adopt a multidimensional set of evaluation metrics to systematically analyze the performance of different syntactic construct configurations. Given the inherent stochasticity of LLM outputs, we conduct $N = 5$ independent runs for each program to ensure the reliability and stability of the experimental results. The specific evaluation metrics are defined as follows: 
\begin{itemize}[leftmargin=*, labelsep=0.5em]
    \item \textbf{Correct Syntactic Construct Compliance Ratio (CSCCR)}: This metric measures the proportion of programs in the dataset for which the generated specifications comply with the constraints imposed by a given configuration, thereby assessing whether an LLM can generate specifications that satisfy the required constraints under given prompts.

    \item \textbf{Number of Verified Programs (NVP)}: This metric quantifies the program verification capability under different syntactic construct configurations. Specifically, it counts the total number of programs for which the generated specifications ultimately cause the verification tool to return success across the five independent runs.
    
    \item \textbf{Number of Stable Verified Programs (NSVP)}: This metric evaluates the robustness of different configurations. To eliminate “lucky” passes caused by the randomness of LLM generation, we define a stably verified program as one that passes verification in at least two out of the five independent runs.

    \item \textbf{Number of Verification Tool Calls (NVTC)}: This metric measures the search cost incurred during the program verification process under different syntactic construct configurations. Specifically, it counts the total number of verification tool calls for a given program within a complete “guess–verify–refine” loop, including the initial specification check as well as all subsequent calls triggered by verification failures.
    \item \textbf{Running Time (RT)}: This metric evaluates the impact of different syntactic construct configurations on the overall efficiency of specification generation and verification. Specifically, it measures the total elapsed time from the start of specification generation for a program to the completion of the verification process.
\end{itemize}
\section{Evaluation Results}
\subsection{RQ1. Feasibility of Logical Constructs}

\begin{figure}[htbp]
  \centering
  
  \begin{minipage}{0.5\linewidth}
    \centering
    \includegraphics[width=\linewidth]{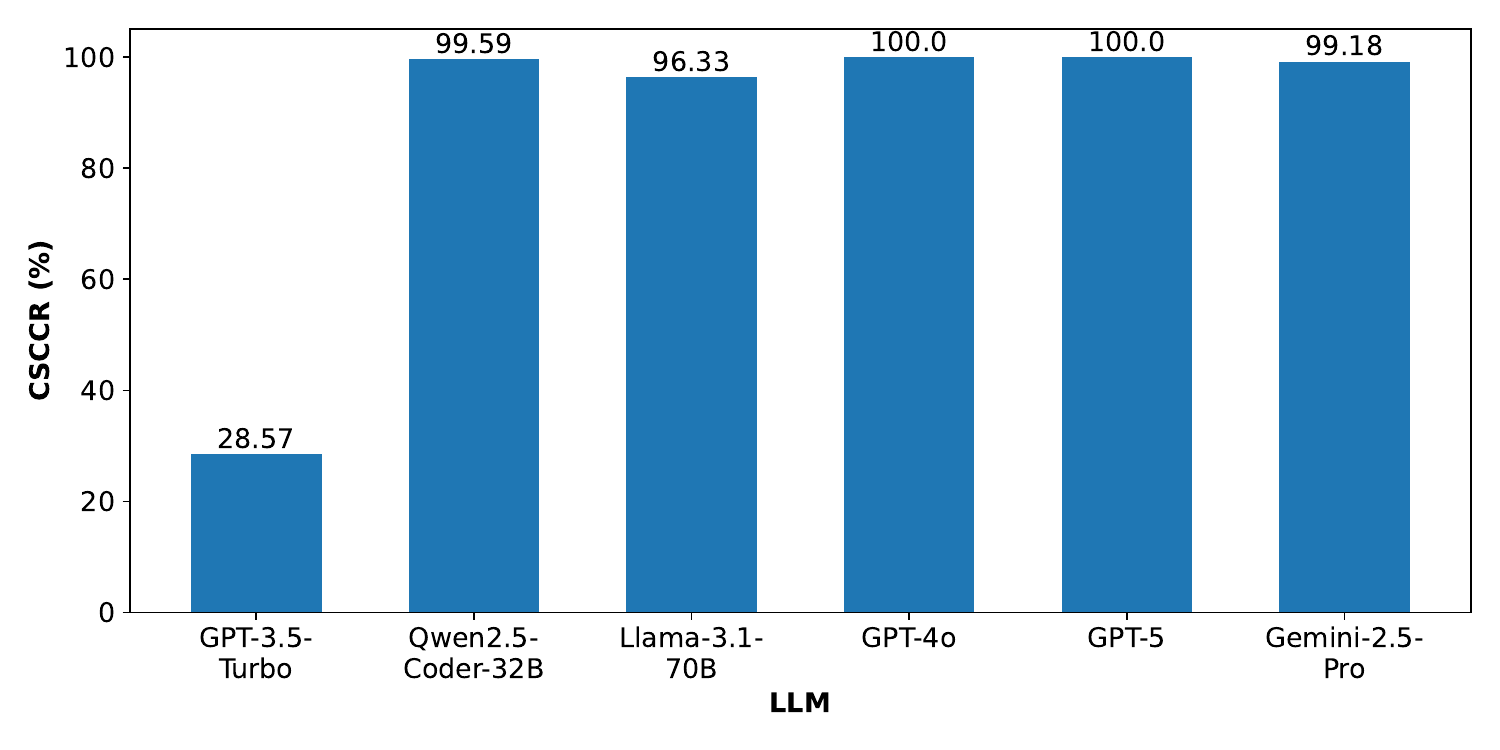}
    \caption*{(a) CSCCR of CV.}
  \end{minipage}%
  \begin{minipage}{0.5\linewidth}
    \centering
    \includegraphics[width=\linewidth]{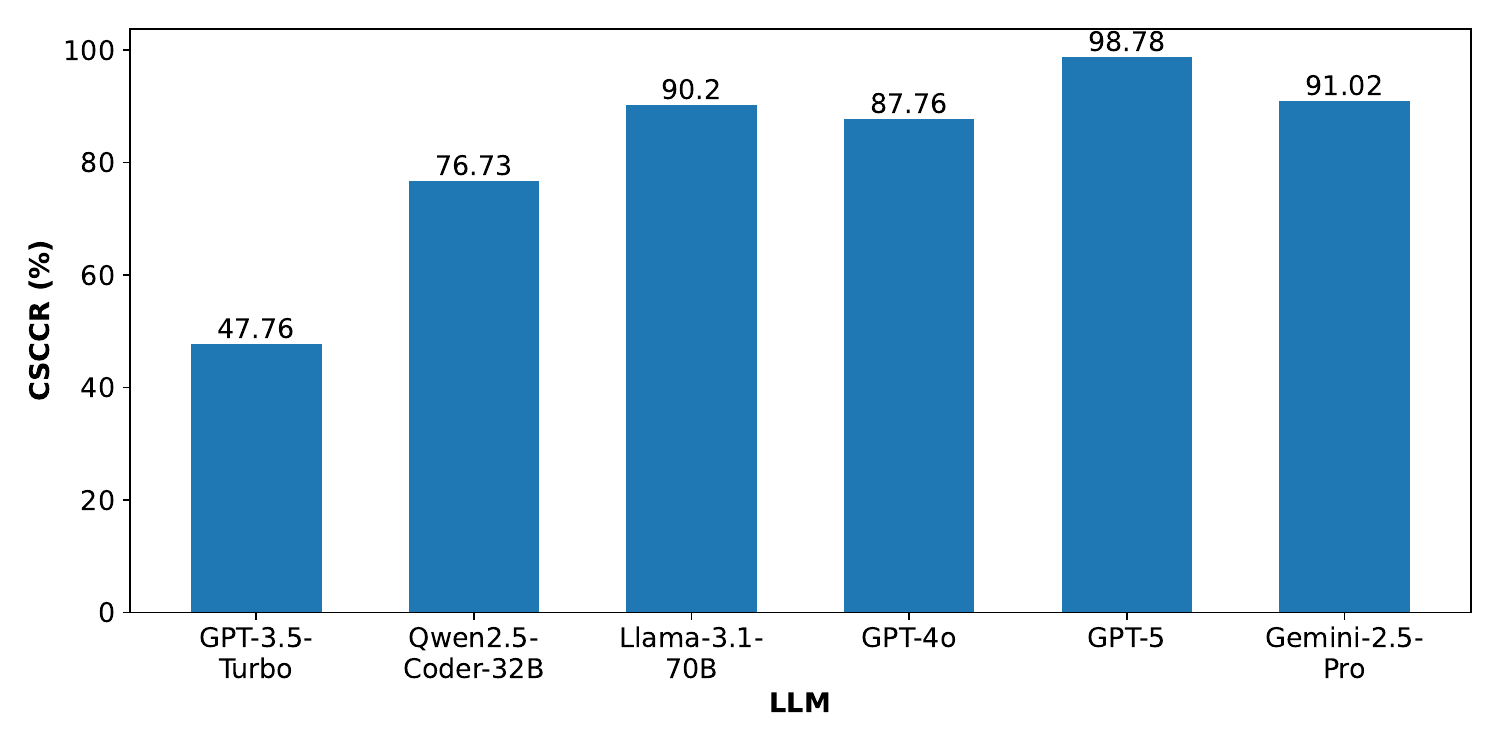}
    \caption*{(b) CSCCR of CA.}
  \end{minipage}

  \begin{minipage}{0.8\linewidth}
    \centering
    \includegraphics[width=\linewidth]{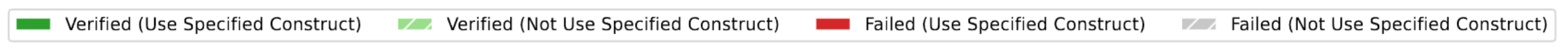}
  \end{minipage}
  \vspace{-0.15em}

  \begin{minipage}{0.5\linewidth}
    \centering
    \includegraphics[width=\linewidth]{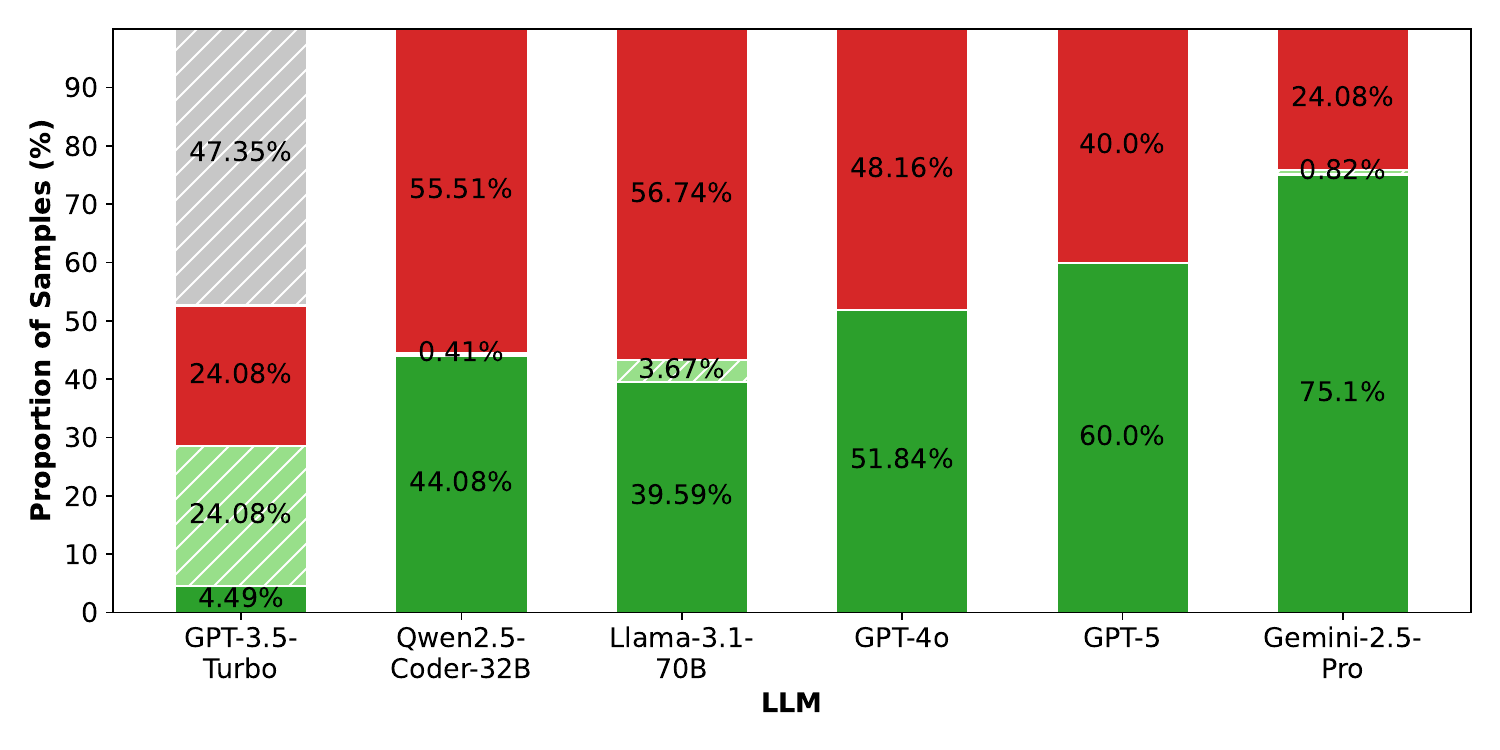}
    \caption*{(c) The Sample Distribution of CV.}
  \end{minipage}%
  \begin{minipage}{0.5\linewidth}
    \centering
    \includegraphics[width=\linewidth]{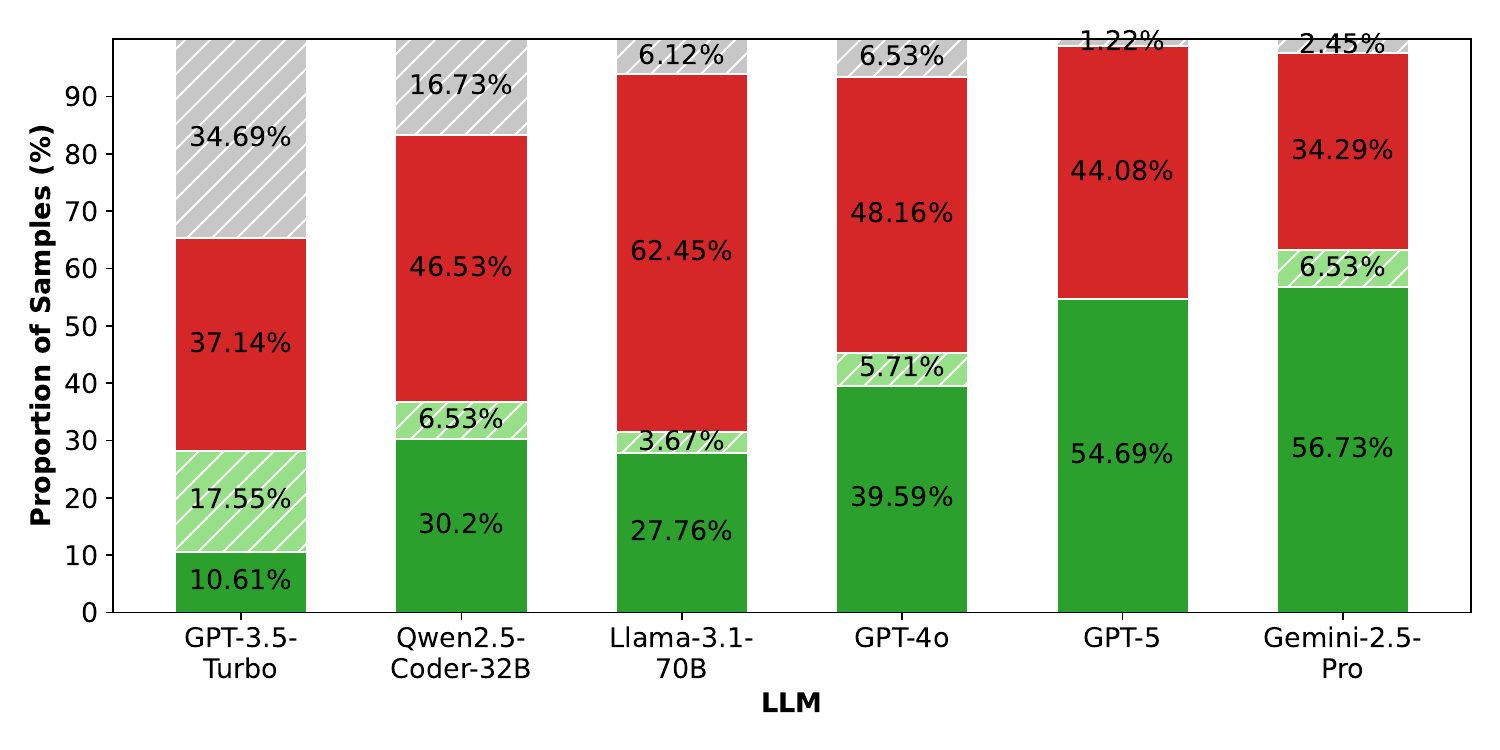}
    \caption*{(d) The Sample Distribution of CA.}
  \end{minipage}

  \caption{The CSCCR and Sample Distribution of Two Logical Construct Configurations.}
  \label{fig: csccr_sd}
\end{figure}

Fig.~\ref{fig: csccr_sd} illustrates the evaluation results of different LLMs’ adherence capability in generating specifications containing the required logical constructs under the deletion paradigm, across two logical construct configurations. Specifically, Fig.~\ref{fig: csccr_sd}-(a) and \ref{fig: csccr_sd}-(b) present the CSCCR metric for the two configurations, respectively, while Fig.~\ref{fig: csccr_sd}-(c) and \ref{fig: csccr_sd}-(d) further depict the distribution of results based on 245 evaluation samples (49 programs $\times$ 5 independent trials).

As shown in Fig.~\ref{fig: csccr_sd}-(a) and \ref{fig: csccr_sd}-(b), GPT‑3.5‑Turbo and Qwen2.5‑Coder‑32B under the CA configuration exhibit substantial gaps in the CSCCR metric compared with the other models, with CSCCR values of 28.57\%, 47.76\%, and 76.73\%, respectively, whereas the remaining models demonstrate consistently high compliance ratios. The sample distributions presented in Fig.~\ref{fig: csccr_sd}-(b) and \ref{fig: csccr_sd}-(d) further help explain this phenomenon. Specifically, for models with lower CSCCR scores, among the samples that fail to employ the specified logical constructs, the proportion of verification failures is significantly higher than that of verification successes. This suggests that once these models generate non‑compliant specifications, limitations in their generative capability result in verification failure for the majority of programs.

It is worth noting that even for more capable models, there remain samples that fail to comply with the configuration requirements, and this phenomenon is more pronounced under the CA configuration. However, a closer examination of the evaluation samples reveals that the underlying causes of non‑compliance under CA differ across models. Specifically, most non‑compliant samples generated by GPT‑3.5‑Turbo rely exclusively on basic syntactic constructs when producing specifications, entirely avoiding the use of logical constructs. In contrast, other models, when failing to adhere to the CA requirements, typically retain verifiable logical constructs such as predicates and logical functions, while omitting only axioms, which represent a more complex syntactic construct. These two forms of non‑compliance reflect distinct failure modes. For less capable models, violations primarily arise from an insufficient understanding of complex logical constructs, leading the model to revert globally to the safest and most conservative set of basic syntactic constructs. For more capable models, non‑compliance more closely resembles a strategic choice: when a model determines that certain programs can be verified using only simpler constructs, it may deliberately omit more complex syntactic constructs, thereby formally deviating from the configuration‑specified syntactic subset. It should be emphasized that, although such behavior reduces the CSCCR metric, it does not necessarily correspond to a failure at the level of logical correctness.


\begin{tcolorbox}[colframe=black, colback=gray!10, coltitle=black, boxrule=0.4mm]
    \textbf{Finding 1:} Introducing logical constructs into the specification generation framework is feasible. With the exception of earlier low‑capacity models, most modern LLMs can comply with configuration requirements and generate specifications using the prescribed logical constructs. Moreover, we identify two fundamentally distinct patterns of logical construct violations: one arises from insufficient model capability, causing the generation process to globally regress to basic syntactic constructs; the other stems from a simplification‑oriented strategic choice made by more capable models during generation.
\end{tcolorbox}

\begin{table}[htbp]
\centering
\caption{Overall Performance.}
\label{table: overall_performance}
\adjustbox{max width=\textwidth}{
\begin{threeparttable}
\begin{tabular}{@{}lcccccccccccccccc@{}}
\toprule
\multicolumn{1}{c|}{\multirow{2}{*}{\textbf{LLMs}}} & \multicolumn{4}{c|}{\textbf{CB}}                                                                          & \multicolumn{4}{c|}{\textbf{CV}}                                                                    & \multicolumn{4}{c|}{\textbf{CA}}                                                                         & \multicolumn{4}{c}{\textbf{CF}}                                                                          \\ \cmidrule(l){2-17} 
\multicolumn{1}{c|}{}                      & \multicolumn{1}{c}{NVP} & \multicolumn{1}{c}{NSVP} & \multicolumn{1}{c}{NVTC} & \multicolumn{1}{c|}{VT} & \multicolumn{1}{c}{NVP} & \multicolumn{1}{c}{NSVP} & \multicolumn{1}{c}{NVTC} & \multicolumn{1}{c|}{VT} & \multicolumn{1}{c}{NVP} & \multicolumn{1}{c}{NSVP} & \multicolumn{1}{c}{NVTC} & \multicolumn{1}{c|}{VT} & \multicolumn{1}{c}{NVP} & \multicolumn{1}{c}{NSVP} & \multicolumn{1}{c}{NVTC} & \multicolumn{1}{c}{VT} \\ \midrule
\multicolumn{17}{c}{\textbf{Deletion Paradigm}}                                                                                                                                                                                                                                                                                                                                                                                                                                                      \\ \midrule
\multicolumn{1}{c|}{GPT-3.5-Turbo}         &             21            &             19             &              116.2             & \multicolumn{1}{c|}{1073.39}    &            19(\textcolor[HTML]{c70040}{$\downarrow\!11$})             &            15(\textcolor[HTML]{c70040}{$\downarrow\!12$})              &              143.8             & \multicolumn{1}{c|}{1374.65}    &          24(\textcolor[HTML]{c70040}{$\downarrow\!8$})               &            17(\textcolor[HTML]{c70040}{$\downarrow\!12$})              &             153.4              & \multicolumn{1}{c|}{1347.90}    &            19             &            18              &            143.8               &            1423.57             \\
\multicolumn{1}{c|}{Qwen2.5-Coder-32B}     &             27            &             24             &              137             & \multicolumn{1}{c|}{889.51}    &            33             &            25              &            146               & \multicolumn{1}{c|}{1128.77}    &          30(\textcolor[HTML]{c70040}{$\downarrow\!4$})               &            20(\textcolor[HTML]{c70040}{$\downarrow\!2$})              &             150.2              & \multicolumn{1}{c|}{1176.14}    &            35             &            27              &              139.2             &            1020.23             \\
\multicolumn{1}{c|}{Llama-3.1-70B}         &             30            &             29             &              128             & \multicolumn{1}{c|}{1149.64}    &            28             &            23(\textcolor[HTML]{c70040}{$\downarrow\!1$})              &               136.6            & \multicolumn{1}{c|}{1645.79}    &          25(\textcolor[HTML]{c70040}{$\downarrow\!1$})               &            21(\textcolor[HTML]{c70040}{$\downarrow\!2$})              &             119.8              & \multicolumn{1}{c|}{1827.20}    &            32             &            28              &            133.8               &             1354.57            \\
\multicolumn{1}{c|}{GPT-4o}                &             28            &             28             &              104.4             & \multicolumn{1}{c|}{845.33}    &            33             &            29              &              149             & \multicolumn{1}{c|}{1205.19}    &          33(\textcolor[HTML]{c70040}{$\downarrow\!1$})               &            27(\textcolor[HTML]{c70040}{$\downarrow\!2$})              &             118              & \multicolumn{1}{c|}{1022.53}    &            35             &            32              &             112.2              &             830.02            \\
\multicolumn{1}{c|}{GPT-5}                 &             32            &             28             &              115.6             & \multicolumn{1}{c|}{757.77}    &            35             &            33              &              132.8             & \multicolumn{1}{c|}{863.29}    &          37               &            29              &              113.6             & \multicolumn{1}{c|}{918.49}    &            40             &            39              &               107.8            &               632.47          \\
\multicolumn{1}{c|}{Gemini-2.5-Pro}        &             42            &             38             &              102.2             & \multicolumn{1}{c|}{2391.11}    &            44             &            40              &               118.2            & \multicolumn{1}{c|}{2310.11}    &          42(\textcolor[HTML]{c70040}{$\downarrow\!3$})               &            37(\textcolor[HTML]{c70040}{$\downarrow\!1$})              &            108.6               & \multicolumn{1}{c|}{2962.98}    &            43             &            42              &             105.8              &            2497.99             \\ \midrule
\multicolumn{1}{c|}{\textbf{Average}}        &             31.8            &             29.4             &              117.44             & \multicolumn{1}{c|}{1206.67}    &            34.6             &            30              &               136.52            & \multicolumn{1}{c|}{1430.63}    &          33.4              &            26.8              &            122.04               & \multicolumn{1}{c|}{1581.47}    &            37             &            33.6              &             119.76              &            1267.06             \\ \midrule
\multicolumn{17}{c}{\textbf{Modification Paradigm}}                                                                                                                                                                                                                                                                                                                                                                                                                                                        \\ \midrule
\multicolumn{1}{c|}{GPT-3.5-Turbo}         &            26             &            23              &              213.4             & \multicolumn{1}{c|}{2595.49}    &            23(\textcolor[HTML]{c70040}{$\downarrow\!16$})             &            20(\textcolor[HTML]{c70040}{$\downarrow\!15$})              &              228.2             & \multicolumn{1}{c|}{3871.61}    &            24(\textcolor[HTML]{c70040}{$\downarrow\!7$})             &             17(\textcolor[HTML]{c70040}{$\downarrow\!12$})             &            239.2               & \multicolumn{1}{c|}{2595.55}    &            22             &              22            &           224.4                &            2152.99             \\
\multicolumn{1}{c|}{Qwen2.5-Coder-32B}     &            31             &            27              &              193.8             & \multicolumn{1}{c|}{1950.18}    &            33             &            29              &             186.2              & \multicolumn{1}{c|}{1978.64}    &            33(\textcolor[HTML]{c70040}{$\downarrow\!5$})             &             24(\textcolor[HTML]{c70040}{$\downarrow\!4$})             &            206.4               & \multicolumn{1}{c|}{2210.20}    &            38             &            29              &           184.8                &            1978.03             \\
\multicolumn{1}{c|}{Llama-3.1-70B}         &            35             &            33              &              168.4             & \multicolumn{1}{c|}{3746.44}    &               35          &                  32        &            169               & \multicolumn{1}{c|}{4576.49}    &          34(\textcolor[HTML]{c70040}{$\downarrow\!2$})               &             32(\textcolor[HTML]{c70040}{$\downarrow\!1$})             &              193.4             & \multicolumn{1}{c|}{5529.16}    &            35             &            32              &              177.4             &           4752.49              \\
\multicolumn{1}{c|}{GPT-4o}                &            35             &            32              &              159.6             & \multicolumn{1}{c|}{1628.86}    &            39             &            37              &             156.4              & \multicolumn{1}{c|}{1526.22}    &            38(\textcolor[HTML]{c70040}{$\downarrow\!1$})             &             34(\textcolor[HTML]{c70040}{$\downarrow\!3$})             &            158               & \multicolumn{1}{c|}{2143.61}    &            35             &            33              &           152.6                &            1486.39             \\
\multicolumn{1}{c|}{GPT-5}                 &            39             &            37              &              144.2             & \multicolumn{1}{c|}{1248.68}    &            40             &            36              &             147.2              & \multicolumn{1}{c|}{1295.14}    &            42             &             41             &            141.2               & \multicolumn{1}{c|}{1533.03}    &            42             &             41             &           116.2                &            1158.36            \\ 
\multicolumn{1}{c|}{Gemini-2.5-Pro}        &            47             &            47              &              102.2             & \multicolumn{1}{c|}{5588.76}    &            48             &            46              &             109              & \multicolumn{1}{c|}{5540.15}    &           47(\textcolor[HTML]{c70040}{$\downarrow\!1$})              &            47(\textcolor[HTML]{c70040}{$\downarrow\!1$})              &              105.4             & \multicolumn{1}{c|}{4798.13}    &            47             &             47             &          103.6                 &           5049.15              \\ \midrule
\multicolumn{1}{c|}{\multirow{1}{*}{\textbf{Average}}}        &             37.4            &             35.2             &              153.64             & \multicolumn{1}{c|}{2832.58}    &            39             &            36              &               153.56            & \multicolumn{1}{c|}{2983.33}    &          38.8              &            35.4              &            160.88               & \multicolumn{1}{c|}{3242.83}    &            39.4             &            36.4              &             146.92              &            2884.88             \\
\multicolumn{1}{c|}{\textbf{Improvement Ratio}}        &             \textcolor[HTML]{008080}{$\uparrow\!17.61\%$}            &             \textcolor[HTML]{008080}{$\uparrow\!19.73\%$}             &              \textcolor[HTML]{008080}{$\uparrow\!30.82\%$}             & \multicolumn{1}{c|}{\textcolor[HTML]{008080}{$\uparrow\!134.74\%$}}    &            \textcolor[HTML]{008080}{$\uparrow\!12.72\%$}             &            \textcolor[HTML]{008080}{$\uparrow\!20.00\%$}              &               \textcolor[HTML]{008080}{$\uparrow\!12.48\%$}            & \multicolumn{1}{c|}{\textcolor[HTML]{008080}{$\uparrow\!108.53\%$}}    &          \textcolor[HTML]{008080}{$\uparrow\!16.17\%$}              &            \textcolor[HTML]{008080}{$\uparrow\!32.09\%$}              &            \textcolor[HTML]{008080}{$\uparrow\!31.83\%$}               & \multicolumn{1}{c|}{\textcolor[HTML]{008080}{$\uparrow\!105.05\%$}}    &            \textcolor[HTML]{008080}{$\uparrow\!6.49\%$}             &            \textcolor[HTML]{008080}{$\uparrow\!8.33\%$}              &             \textcolor[HTML]{008080}{$\uparrow\!22.68\%$}              &            \textcolor[HTML]{008080}{$\uparrow\!127.68\%$}             \\ \bottomrule
\end{tabular}
\end{threeparttable}
}
\end{table}

\subsection{RQ2. Performance Comparison}
Table~\ref{table: overall_performance} summarizes the overall performance of different LLMs under four syntactic construct configurations and two refinement paradigms. The red numbers in the table denote the number of programs that pass verification despite not using the specified logical constructs. As discussed in RQ1, compared with other models, GPT‑3.5‑Turbo produces a large number of non‑compliant samples under both logical construct configurations, which renders its NVP and NSVP metrics incomparable. Consequently, we exclude this model from subsequent performance‑comparison analyses.

\subsubsection{RQ2.1 Comparison of Verification Capability}

\begin{figure}[htbp]
  \centering
  
  \begin{minipage}{0.5\linewidth}
    \centering
    \includegraphics[width=\linewidth]{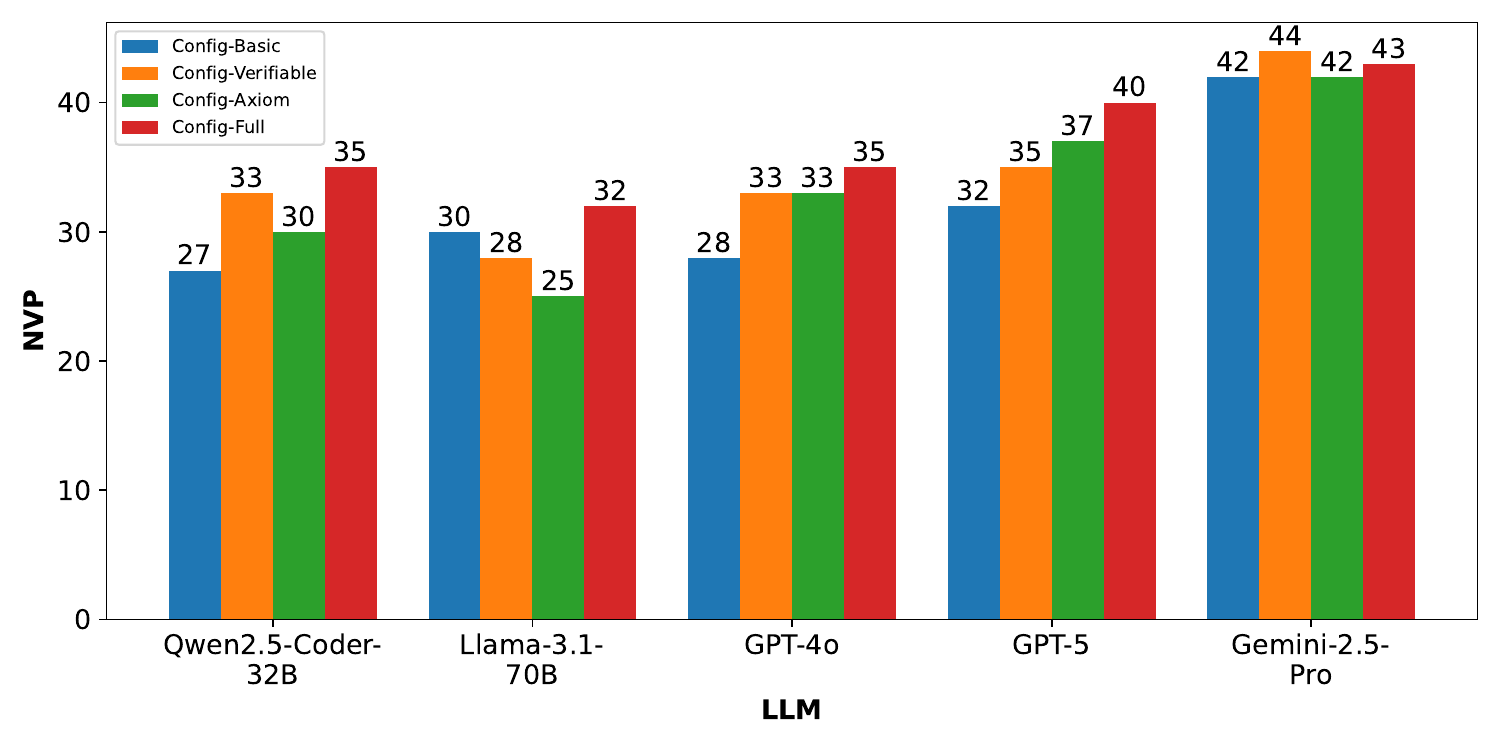}
    \caption*{(a) NVP under Deletion Paradigm.}
  \end{minipage}%
  \begin{minipage}{0.5\linewidth}
    \centering
    \includegraphics[width=\linewidth]{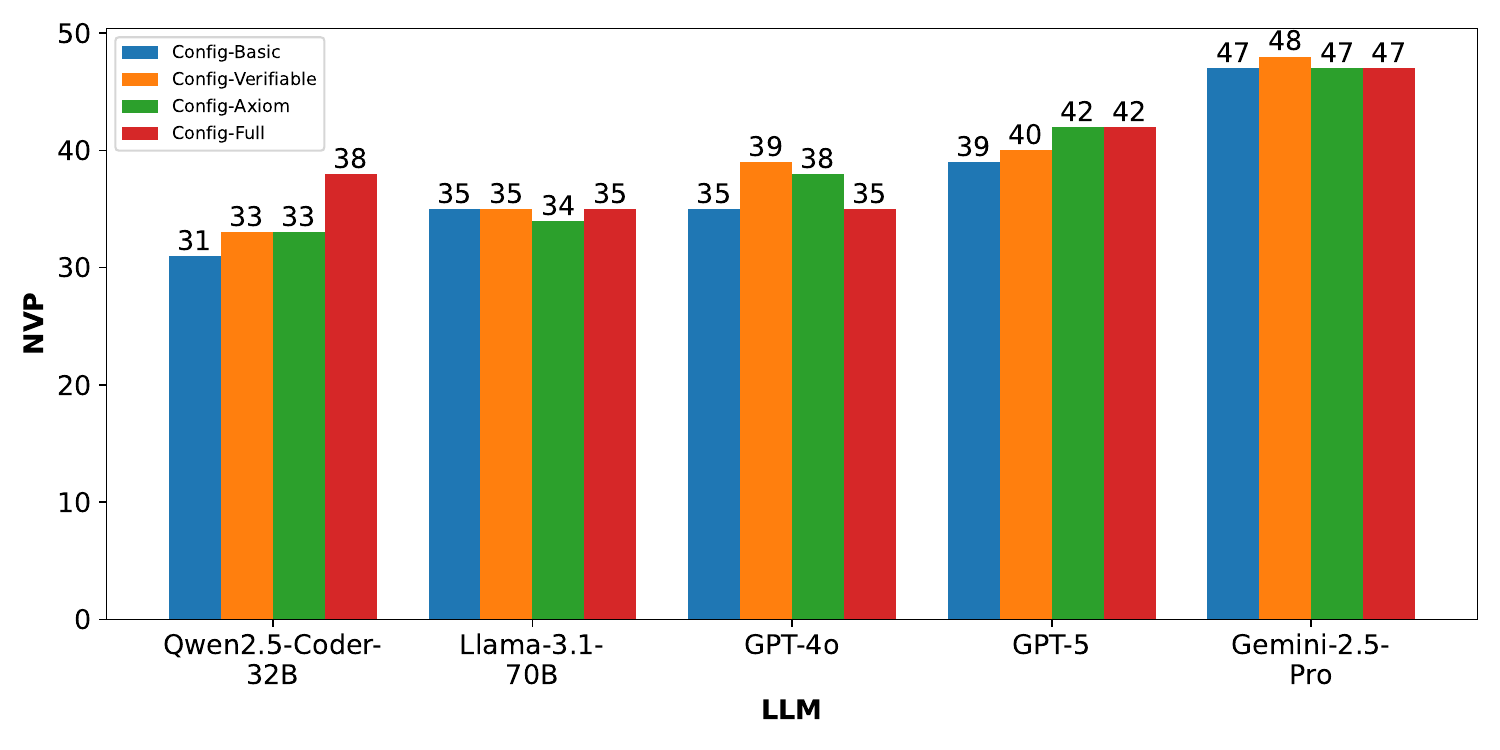}
    \caption*{(b) NVP under Modification Paradigm.}
  \end{minipage}%
  
  \caption{NVP of different LLMs under four syntactic construct configurations and two refinement paradigms.}
  \label{fig: NVP}
\end{figure}

Fig.~\ref{fig: NVP} present the NVP metric for different LLMs under four syntactic construct configurations and two refinement paradigms. Overall, introducing logical constructs substantially improves program verifiability. Compared with CB, the two configurations that employ only the corresponding logical constructs—CV and CA—verify, on average, 8.81\% and 5.03\% more programs under the deletion paradigm, and achieve improvements of 4.28\% and 3.74\%, respectively, under the modification paradigm. Furthermore, CF, which integrates multiple syntactic constructs, demonstrates the strongest verification capability, verifying 16.35\% and 5.35\% more programs than CB under the two refinement paradigms, respectively.

By comparing the performance of the same LLM across different configurations, we observe that for most models, the NVP under CV and CA exceeds that under CB, with Llama‑3.1‑70B under CV being the only exception. However, this observation does not imply that introducing logical constructs weakens verification capability. Rather, the underlying reason is that this model already attains exceptionally strong performance under CB: its NVP not only surpasses that of GPT‑4o, but also approaches the level of GPT‑5. Further examination of the CF results shows that, across all models and both refinement paradigms, this configuration consistently achieves the highest or near‑highest NVP. This suggests that when all syntactic constructs are permitted simultaneously, models benefit from a larger strategic space, enabling coverage of a broader set of verifiable programs. From a cross‑model perspective, as model capability increases, the NVP across all configurations exhibits an overall upward trend, reflecting the substantial impact of intrinsic model capacity on program verification performance. More importantly, the relative gains introduced by logical constructs remain consistently observable across different model scales.

\begin{figure}[htbp]
  \centering

  \begin{minipage}{0.45\linewidth}
    \centering
    \includegraphics[width=\linewidth]{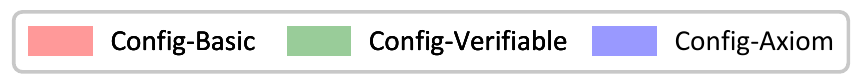}
  \end{minipage}
  \vspace{-0.25em}
  
  \begin{minipage}{0.2\linewidth}
    \centering
    \includegraphics[width=\linewidth]{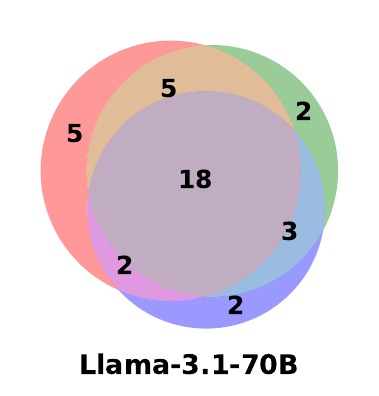}
  \end{minipage}%
  \begin{minipage}{0.2\linewidth}
    \centering
    \includegraphics[width=\linewidth]{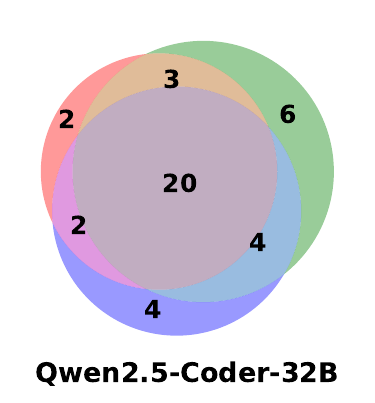}
  \end{minipage}%
  \begin{minipage}{0.2\linewidth}
    \centering
    \includegraphics[width=\linewidth]{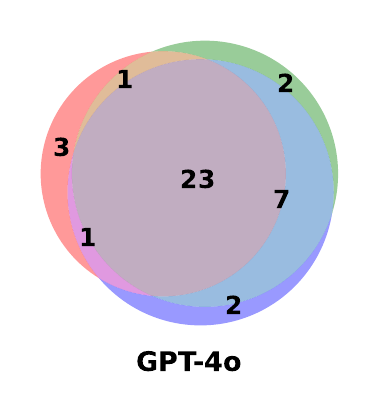}
  \end{minipage}%
  \begin{minipage}{0.2\linewidth}
    \centering
    \includegraphics[width=\linewidth]{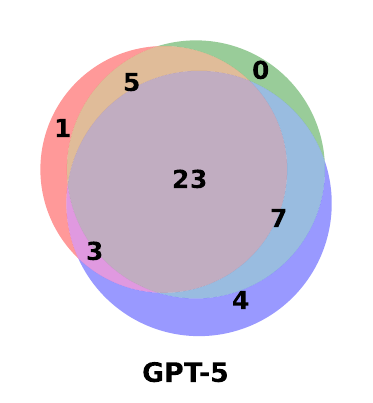}
  \end{minipage}%
  \begin{minipage}{0.2\linewidth}
    \centering
    \includegraphics[width=\linewidth]{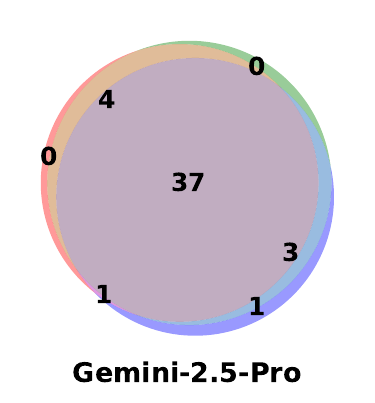}
  \end{minipage}
  
  \caption{Venn Diagrams of the Programs that Passed Verification under the three configurations (CB, CV, and CA) for different LLMs.}
  \label{fig: Venn}
\end{figure}

Furthermore, Fig.~\ref{fig: Venn} visualizes, in the form of Venn diagrams, the relationships among the sets of programs successfully verified under CB, CV, and CA for different LLMs under the deletion paradigm. Each Venn diagram characterizes, for a given model, the subsets of verifiable programs covered by different syntactic construct configurations and their overlaps. Specifically, across all models, we observe program sets that cannot be verified under one configuration but can be verified under others, indicating that specifications generated using a single class of syntactic constructs suffer from inherent limitations in expressiveness. Consequently, the three configurations do not exhibit a simple containment relationship; instead, they demonstrate substantial complementarity in verification capability. From a cross‑model perspective, a clear evolutionary trend emerges as model capability increases. For less capable models, the differences among the three configurations are pronounced, whereas as model capability improves, the intersections among the program sets corresponding to the three configurations gradually expand. This suggests that for low‑capacity models, introducing more appropriate syntactic constructs can substantially compensate for intrinsic model limitations and constitutes a key factor in improving verification performance. In contrast, for high‑capacity models, syntactic constructs primarily reflect differences in expressiveness and modeling style; although their marginal gains are relatively diminished, they remain indispensable for covering specific subsets of programs.

\begin{tcolorbox}[colframe=black, colback=gray!10, coltitle=black, boxrule=0.4mm]
    \textbf{Finding 2:} The introduction of logical declarations significantly enhances the verification capability of the specification generation framework. However, different syntactic constructs do not form a strict hierarchy; instead, they exhibit complementary properties. Consequently, logical constructs cannot fully replace basic syntactic constructs. Moreover, the best verification performance is typically achieved by jointly employing multiple syntactic constructs.
\end{tcolorbox}

\subsubsection{RQ2.2 Comparison of Stable Verification Capabilities}
\begin{figure}[htbp]
  \centering
  
  \begin{minipage}{0.5\linewidth}
    \centering
    \includegraphics[width=\linewidth]{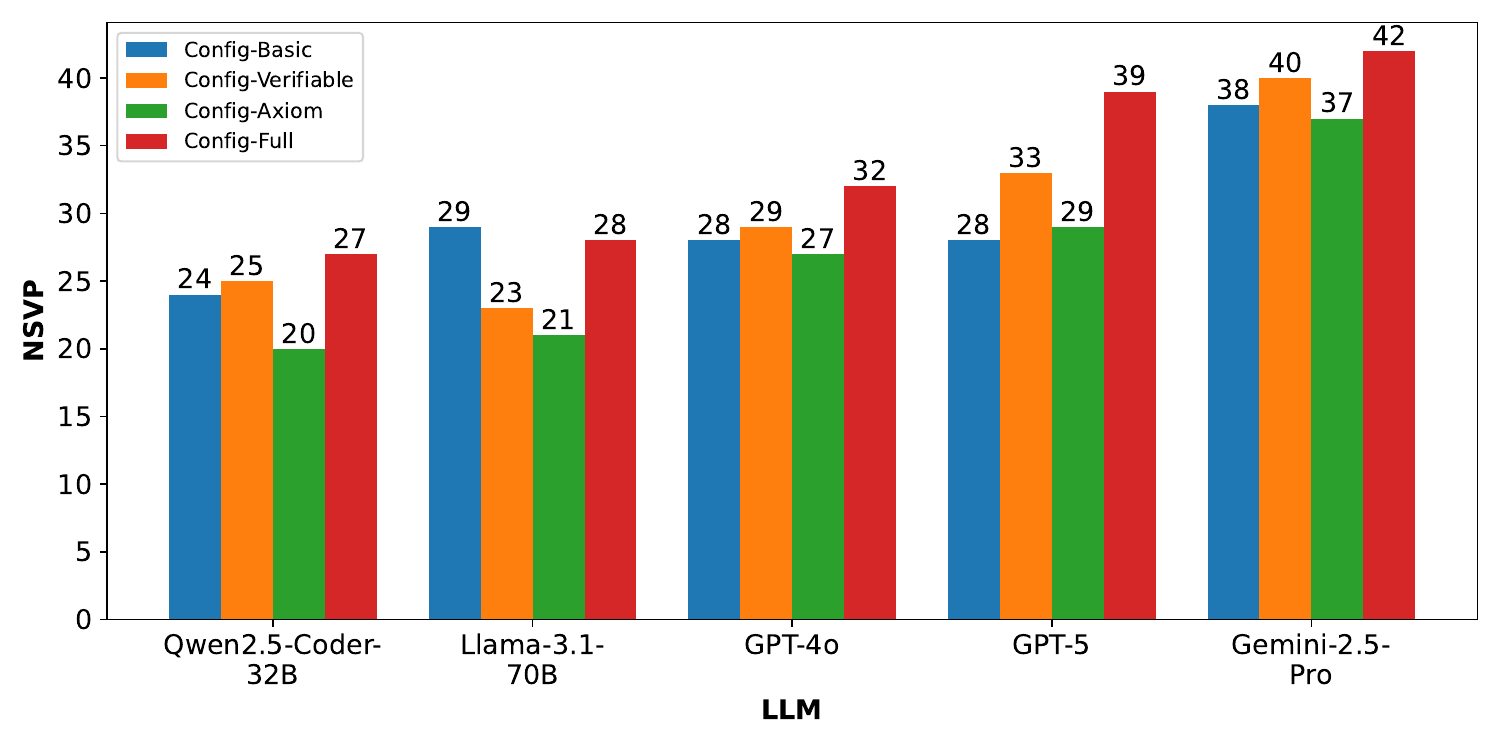}
    \caption*{(a) NSVP under Deletion Paradigm.}
  \end{minipage}%
  \begin{minipage}{0.5\linewidth}
    \centering
    \includegraphics[width=\linewidth]{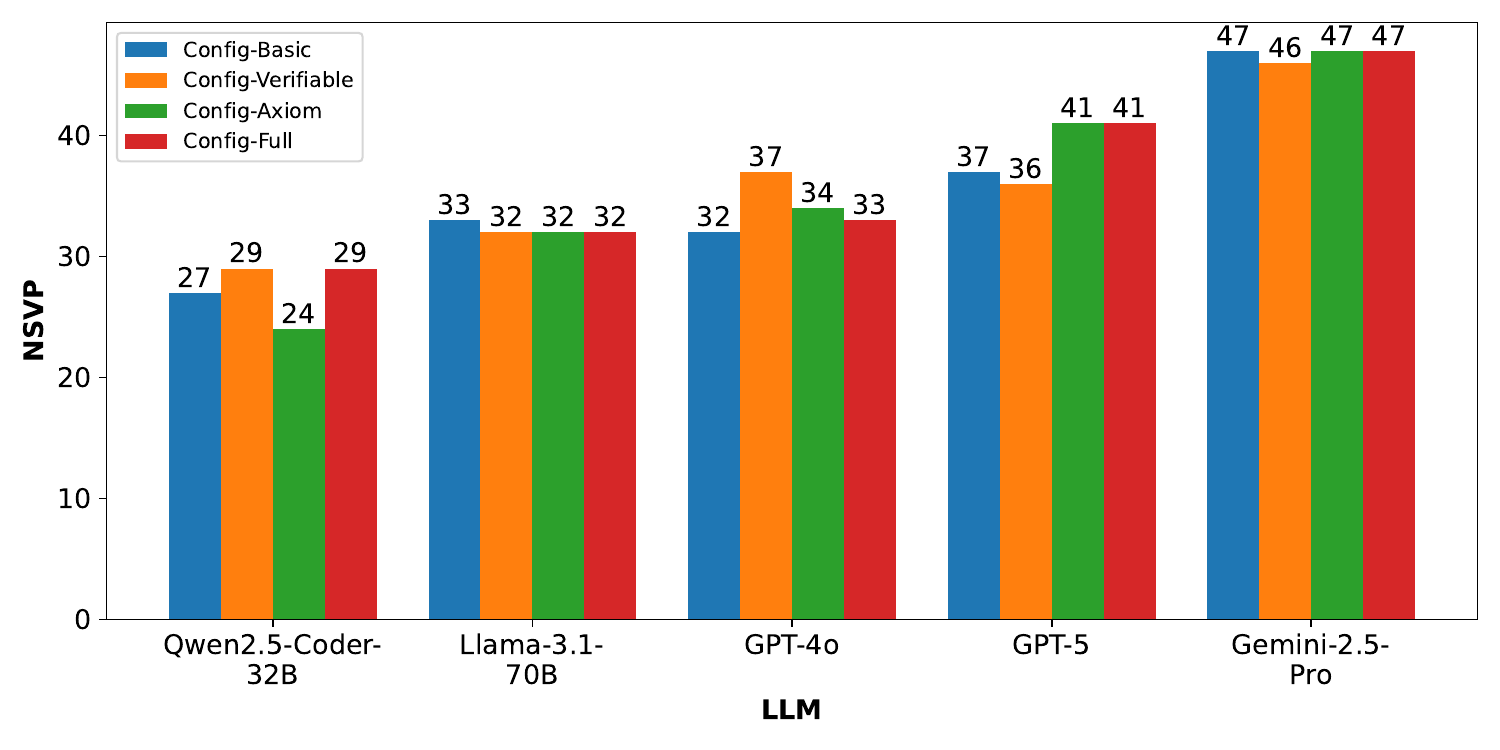}
    \caption*{(b) NSVP under Modification Paradigm.}
  \end{minipage}%

  \begin{minipage}{0.5\linewidth}
    \centering
    \includegraphics[width=\linewidth]{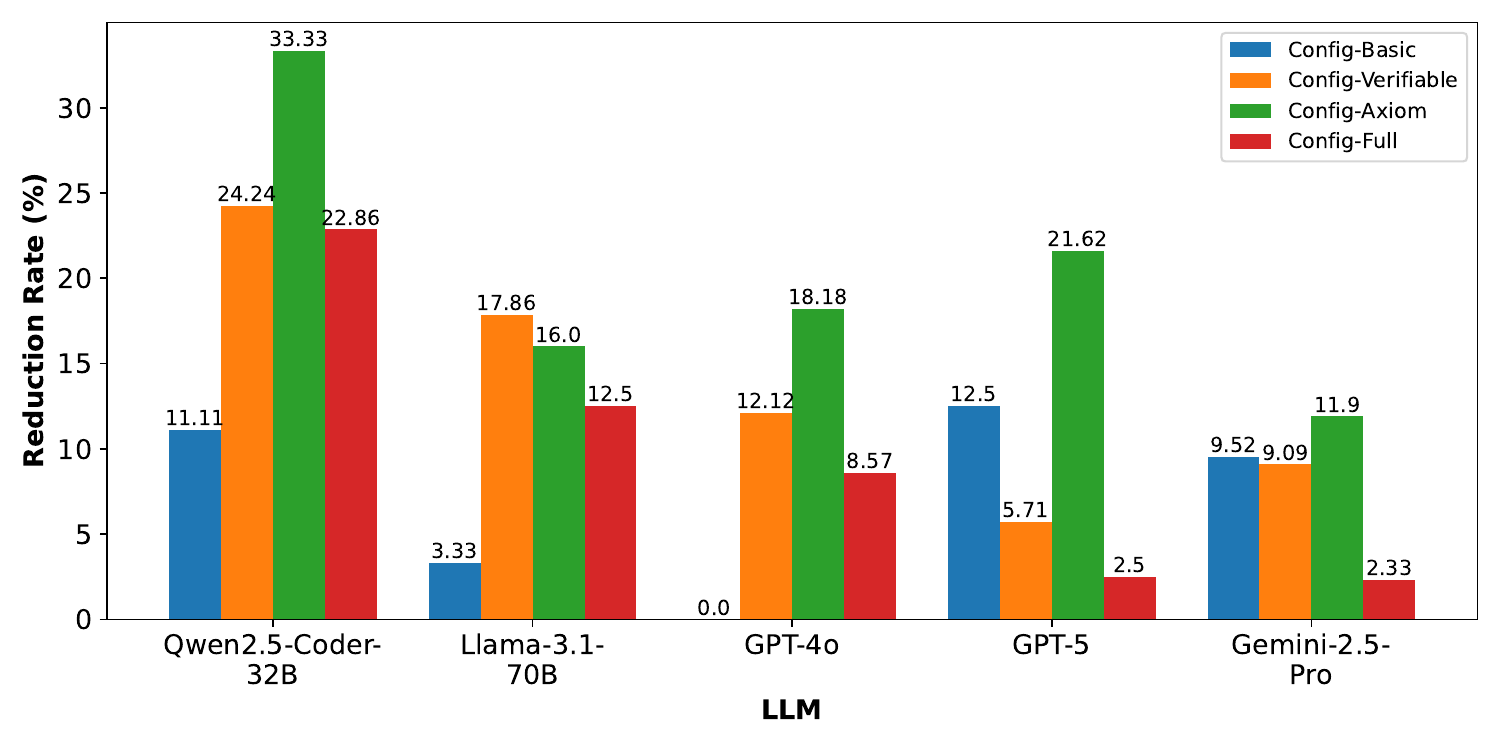}
    \caption*{(c) The Reduction Rate under Deletion Paradigm.}
  \end{minipage}%
  \begin{minipage}{0.5\linewidth}
    \centering
    \includegraphics[width=\linewidth]{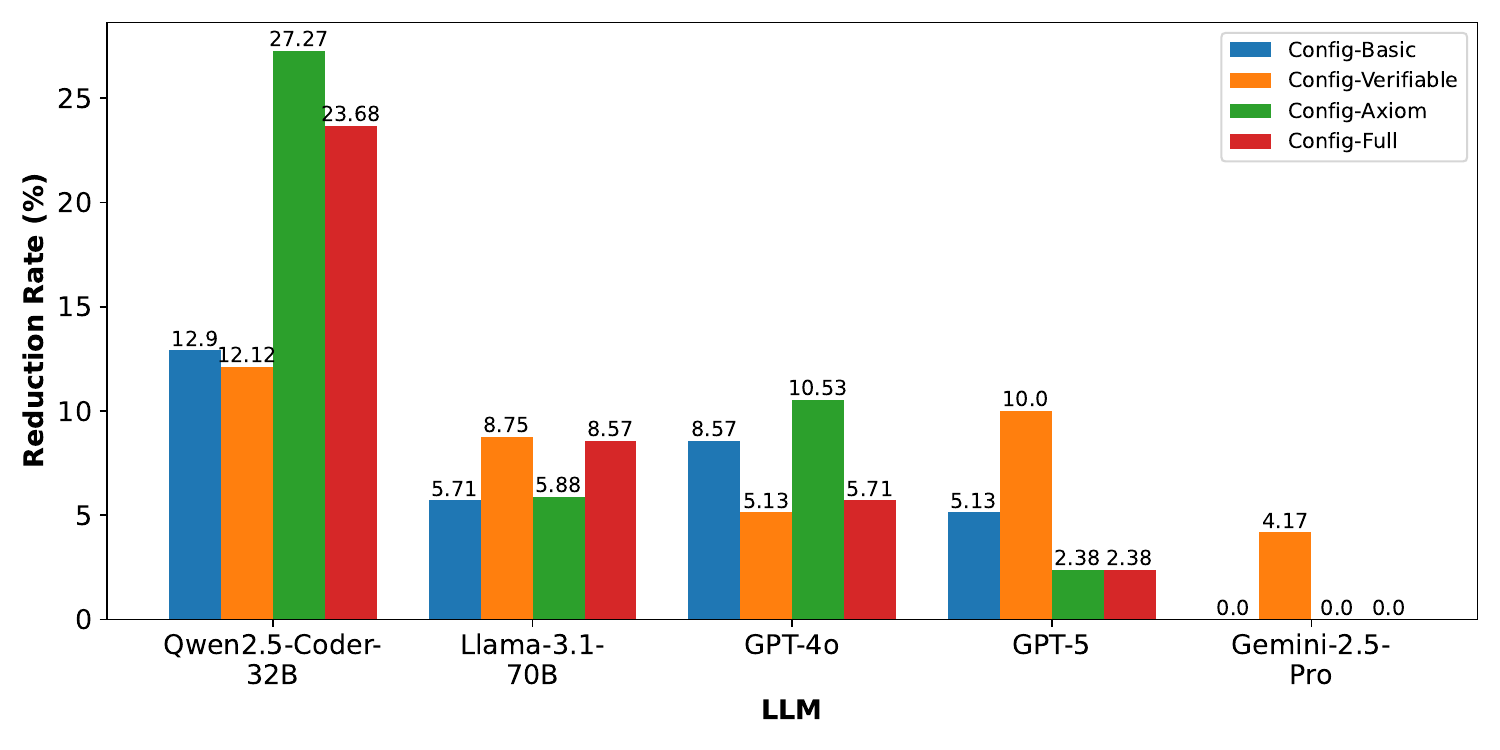}
    \caption*{(d) The Reduction Rate under Modification Paradigm.}
  \end{minipage}%
  
  \caption{Stability Performance of Different LLMs under Four Syntactic Construct Configurations and Two Refinement Paradigms.}
  \label{fig: stability_performance}
\end{figure}

Fig.~\ref{fig: stability_performance} examines stability by reporting the NSVP metric for different LLMs under four syntactic construct configurations and two refinement paradigms, together with the relative reduction of this metric compared to NVP. In conjunction with the data in Table.~\ref{table: overall_performance}, we observe that after excluding programs that cannot be stably verified (i.e., NVP $-$ NSVP), the number of verifiable programs under the deletion paradigm decreases by 7.55\%, 13.29\%, 19.76\%, and 9.19\% for the four configurations, respectively. Under the modification paradigm, the corresponding reductions are 5.88\%, 7.69\%, 8.76\%, and 7.61\%. Taken together with Finding 1, these results suggest that while the introduction of logical constructs improves program verifiability, the resulting gains remain partially constrained with respect to stability.

Specifically, as shown in Fig.~\ref{fig: stability_performance}-(a) and \ref{fig: stability_performance}-(c), under the deletion paradigm, CA exhibits the highest reduction rate for most models, which in turn results in the lowest NSVP among the four configurations. This observation indicates that relying on unverifiable axiom constructs is more prone to introducing instability during specification generation. In contrast, CB consistently maintains a relatively stable reduction rate across all models. This suggests that specifications generated solely using basic syntactic constructs, although limited in expressiveness, tend to have simpler structures and a smaller search space, making it easier for LLMs to maintain consistent generation patterns across multiple independent runs. As model capability increases, the reduction rate of CV gradually decreases, indicating that verifiable logical constructs are better suited to higher‑capacity models. Leveraging the combined strengths of basic syntactic constructs and verifiable logical constructs, CF demonstrates comparatively robust stability across all models and even emerges as the most stable configuration in terms of verification performance for high‑capacity models.

As shown in Fig.~\ref{fig: stability_performance}-(d), under the modification paradigm, with the exception of Qwen2.5‑Coder‑32B, most models exhibit only minor differences in program reduction rates across configurations, and no clear systematic pattern emerges. This suggests that, under this paradigm, the transition from “verifiable” to “stably verifiable” is largely insensitive to configuration choice, such that the NSVP ranking of a given model across configurations essentially inherits its performance ordering under NVP. This phenomenon is directly reflected in Fig.~\ref{fig: stability_performance}-(b): the relative relationships among NSVP values under different configurations closely mirror those of the NVP results in Fig.~\ref{fig: NVP}-(b). These observations indicate that the progressive modification of specifications effectively suppresses the randomness introduced by logical constructs, thereby making verification outcomes more reproducible across repeated experiments.


It is worth noting that under both refinement paradigms, Qwen2.5‑Coder‑32B exhibits an especially pronounced drop in the CA configuration. In light of the overall behavior observed for the other models, a more plausible explanation is that this effect arises from the model’s limited ability to generate and manipulate complex axiom constructs. In other words, when a model is not yet capable of stably handling logical constructs, introducing such constructs can indeed amplify randomness in the generation process.

\begin{tcolorbox}[colframe=black, colback=gray!10, coltitle=black, boxrule=0.4mm]
    \textbf{Finding 3:} Although logical constructs enhance verification capability, they often introduce increased instability, particularly when axiom constructs are employed under the deletion paradigm. However, compared to relying solely on basic syntactic constructs—which consistently exhibit stable behavior—combining basic syntactic constructs with logical constructs yields more stable overall performance on high‑capacity models.
\end{tcolorbox}

\subsubsection{RQ2.3 Comparison of Efficiency}

In this RQ, we adopt NVTC and RT as the core evaluation metrics to systematically analyze the impact of different types of syntactic constructs on verification tool and overall execution cost. For both metrics, we apply the same analysis procedure. First, we compare the metrics across the four configurations to characterize their overall effects. Subsequently, at the program level, we identify—among the CB, CV, and CA configurations—the configuration that yields the fewest verification calls or the shortest runtime, and then compute the proportion of programs for which each configuration emerges as the optimal choice. 

\begin{figure}[htbp]
  \centering

  \begin{minipage}{0.5\linewidth}
    \centering
    \includegraphics[width=\linewidth]{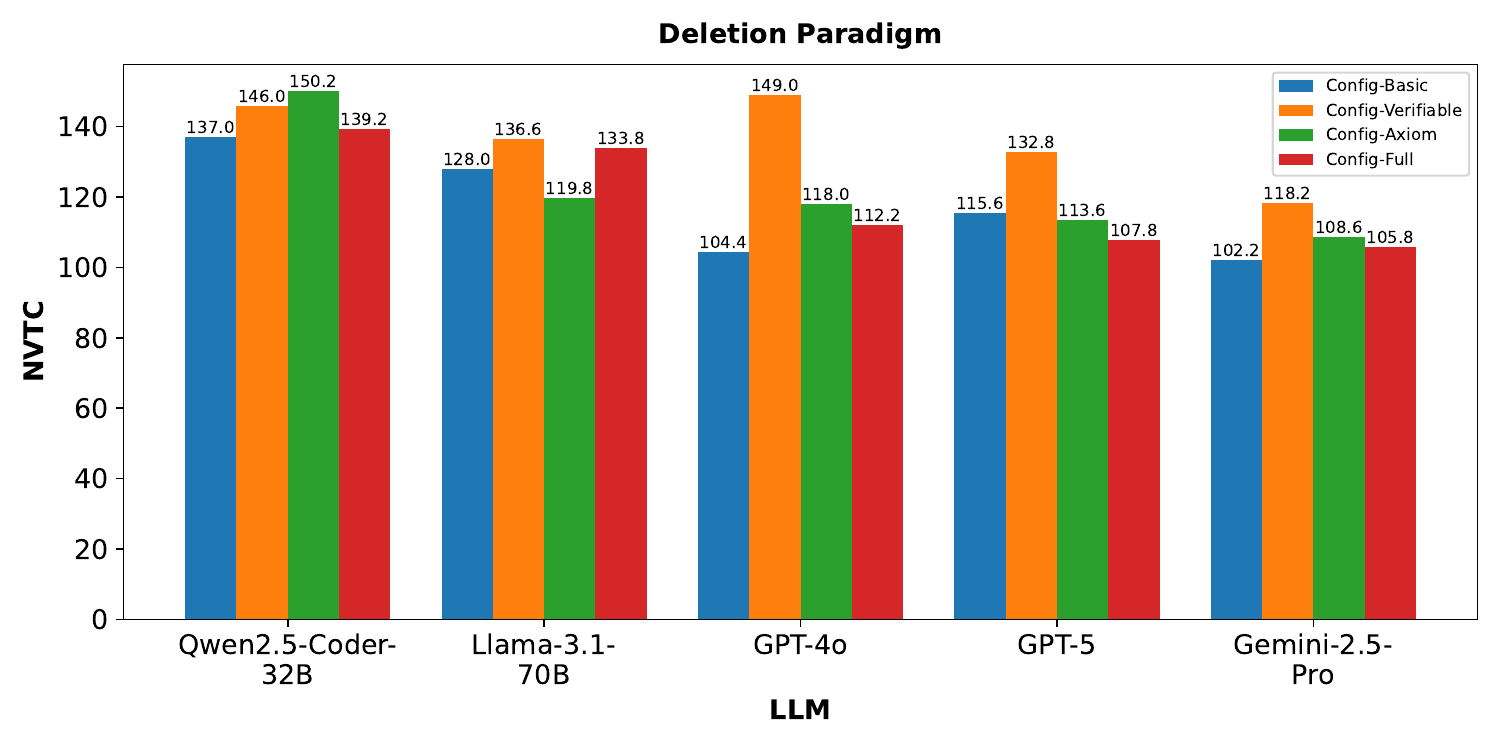}
  \end{minipage}%
  \begin{minipage}{0.5\linewidth}
    \centering
    \includegraphics[width=\linewidth]{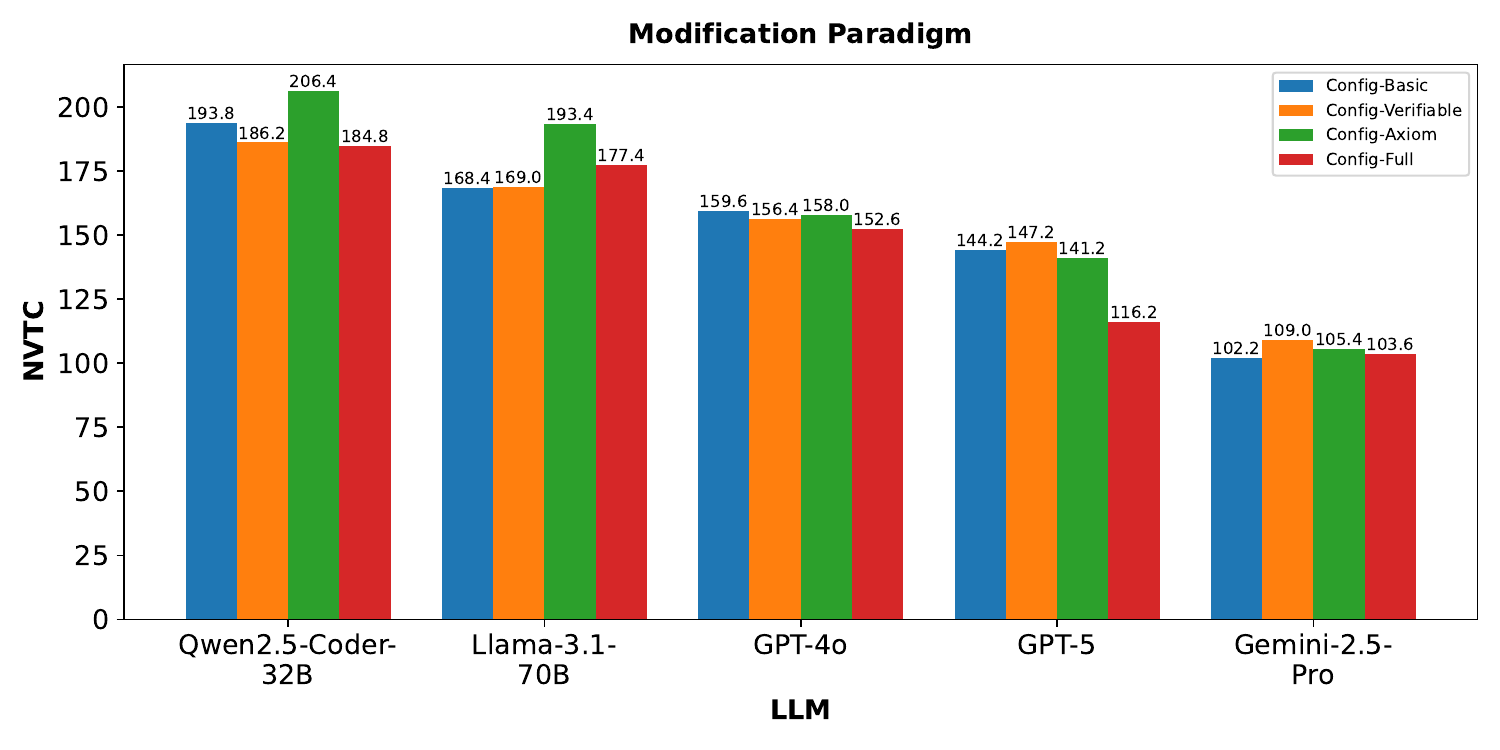}
  \end{minipage}%
  
  \caption{NVTC under Deletion and Modification Paradigm.}
  \label{fig: NVTC}
\end{figure}

In analyzing the impact on the verification tool, at an aggregate level, we observe that under the deletion paradigm, the NVTC of the CV, CA, and CF configurations increases by 16.25\%, 3.91\%, and 1.98\%, respectively, compared with the CB configuration. Combined with the relative relationships among configurations under the same LLM shown in Fig~\ref{fig: NVTC}, these results suggest that while logical constructs enhance expressiveness, they also increase verification complexity, leading the verification tool to undergo more failed attempts and pruning steps. This effect is particularly pronounced for verifiable logical constructs, whereas axiom constructs, due to their non‑verifiable nature, partially mitigate this issue. Notably, across all models, CF performs comparably to CB, indicating that the coordinated use of basic syntactic constructs and logical constructs can improve verification capability without significantly increasing exploration cost of the verification tool. Under the modification paradigm, compared with CB, the NVTC of CA increases by 4.71\%, while that of CF decreases by 4.73\%, and CV remains largely comparable to CB. This behavior can be primarily attributed to the fact that, in the modification paradigm, we impose an upper bound on the maximum number of specification refinement iterations, thereby explicitly constraining the number of verification tool calls that a single program may trigger. As a result, the NVTC differences among the four configurations are substantially reduced relative to those observed under the deletion paradigm.

\begin{figure}[htbp]
  \centering
  
  \begin{minipage}{0.35\linewidth}
    \centering
    \includegraphics[width=\linewidth]{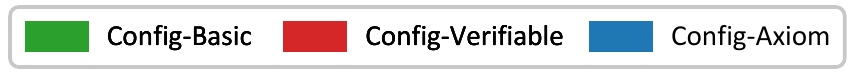}
  \end{minipage}
  \vspace{-0.2em}

  \begin{minipage}{0.5\linewidth}
    \centering
    \includegraphics[width=\linewidth]{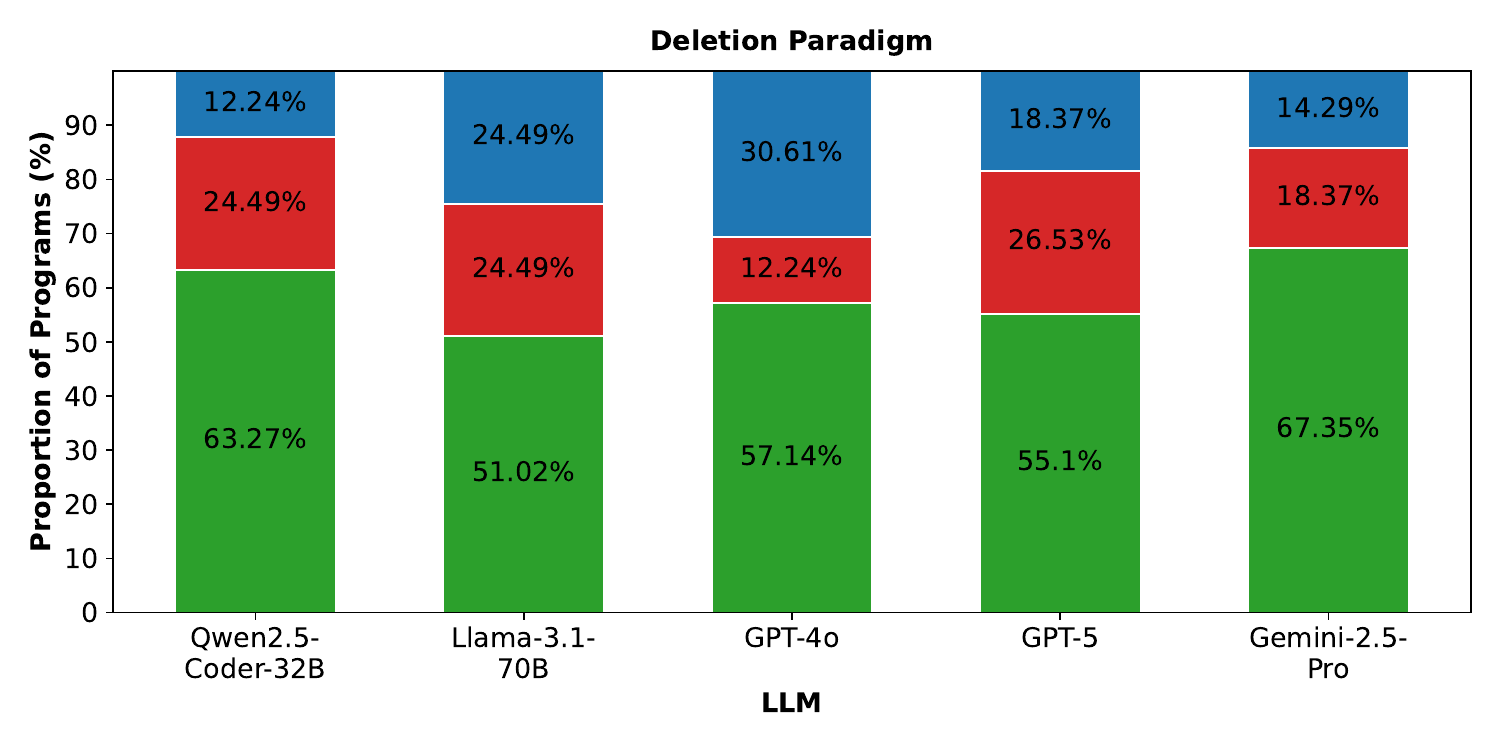}
  \end{minipage}%
  \begin{minipage}{0.5\linewidth}
    \centering
    \includegraphics[width=\linewidth]{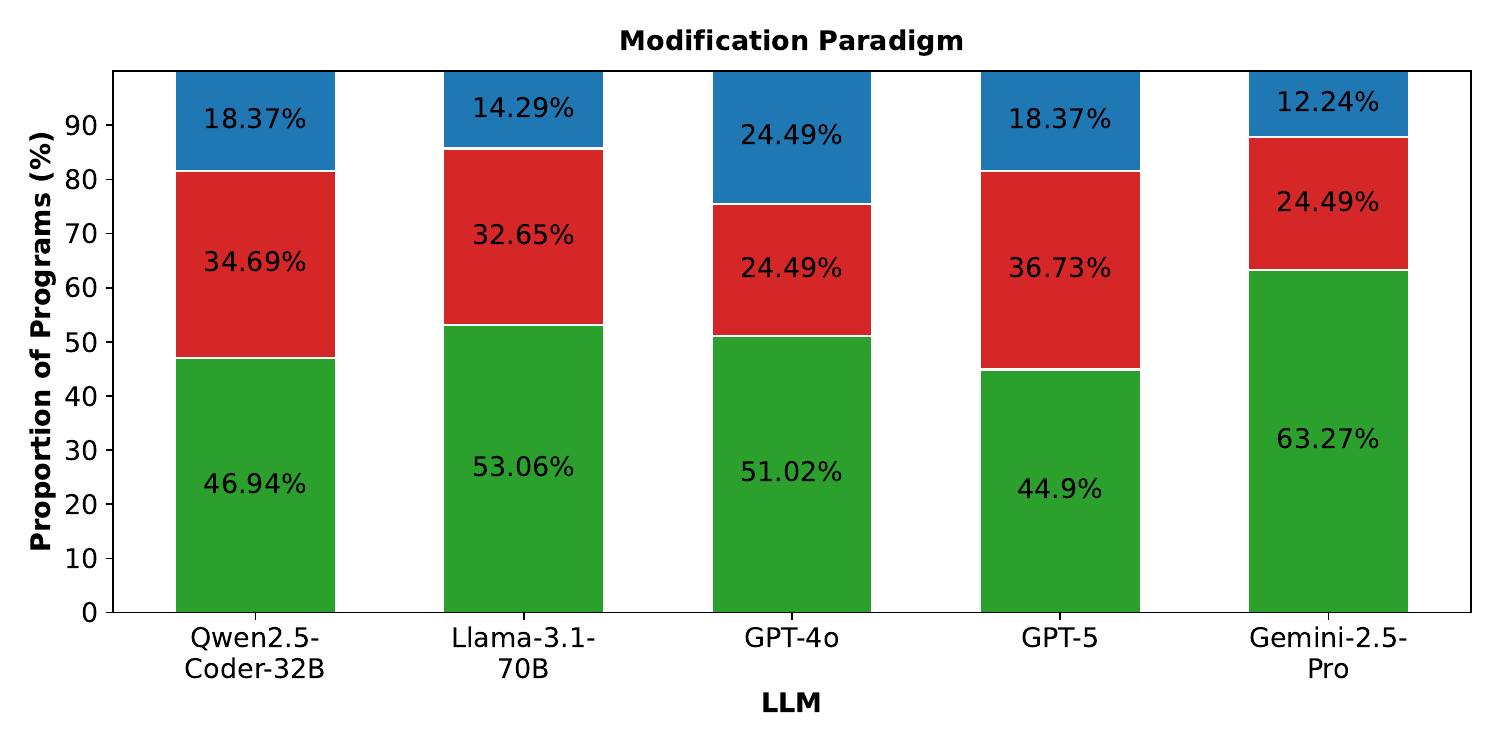}
  \end{minipage}%
  
  \caption{Proportion of Programs with Optimal Configuration (Achieving Minimum NVTC).}
  \label{fig: Proportion_NVCT)}
\end{figure}

Furthermore, Fig.~\ref{fig: Proportion_NVCT)} illustrates the proportion of programs for which the CB, CV, and CA configurations emerge as the optimal choice. Specifically, CB accounts for the highest proportion across all models and, in most cases, yields the lowest NVTC for more than half of the programs. This result indicates that specifications relying solely on basic syntactic constructs can complete verification with fewer verification tool calls for the majority of programs, reflecting advantages in terms of simpler structure and more direct verification paths. Nevertheless, the CV and CA configurations also constitute non‑negligible proportions across all models, suggesting that different configurations exhibit distinct advantage regimes with respect to the NVTC metric.

\begin{figure}[htbp]
  \centering

  \begin{minipage}{0.5\linewidth}
    \centering
    \includegraphics[width=\linewidth]{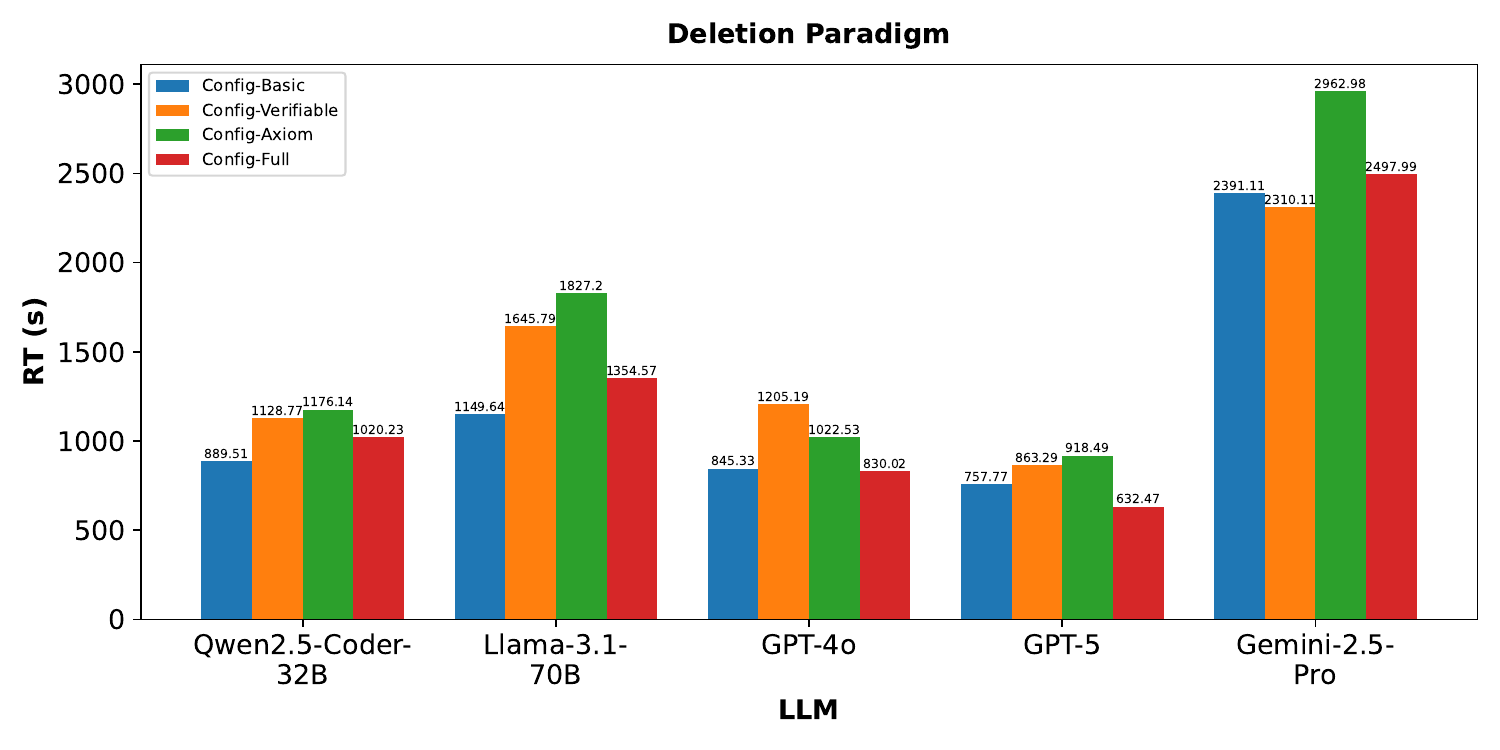}
  \end{minipage}%
  \begin{minipage}{0.5\linewidth}
    \centering
    \includegraphics[width=\linewidth]{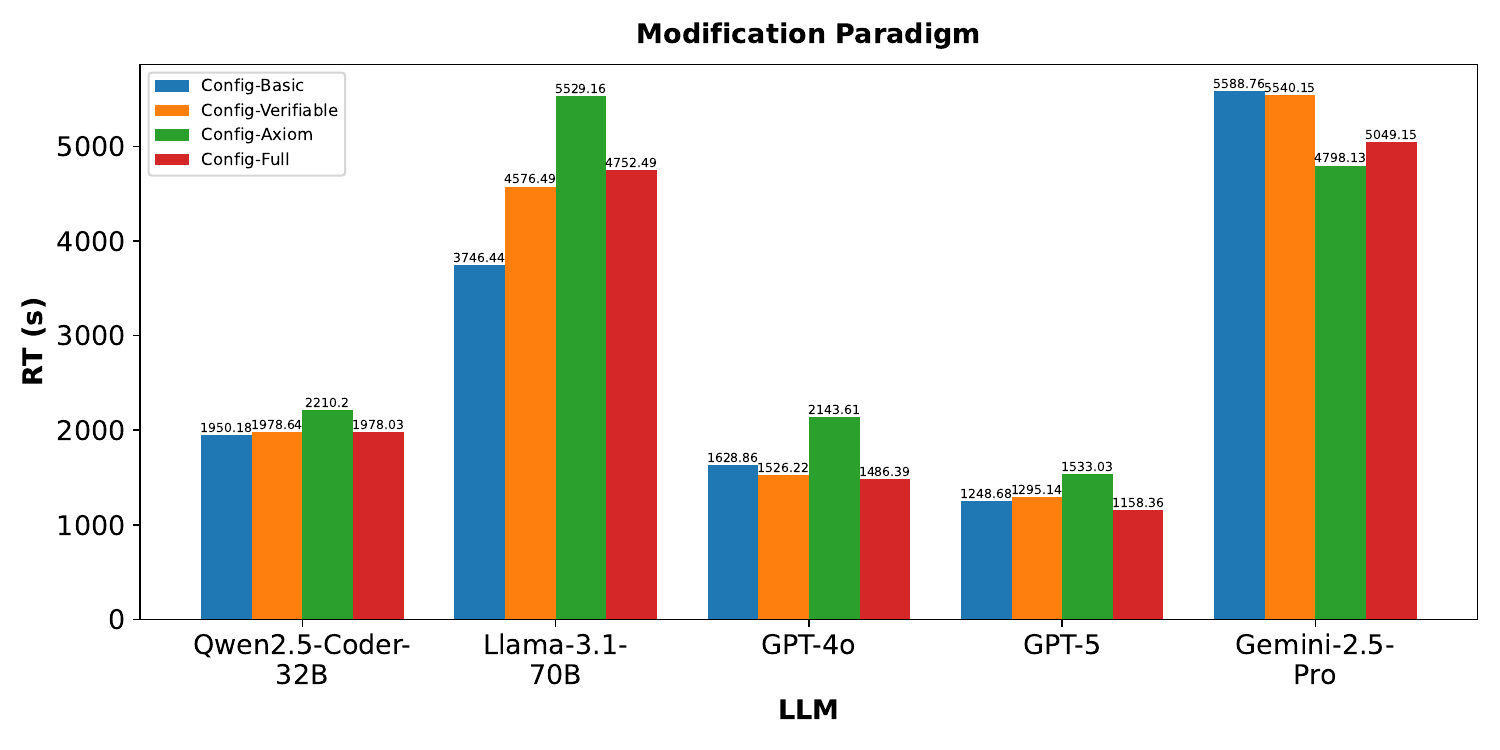}
  \end{minipage}%
  
  \caption{RT under Deletion and Modification Paradigm.}
  \label{fig: RT}
\end{figure}

In analyzing the impact on overall execution cost, at an aggregate level, the RT of the CV, CA, and CF configurations is consistently higher than that of CB. Under the deletion paradigm, the runtime increases for these three configurations are 18.56\%, 31.06\%, and 5.00\%, respectively. As indicated by the detailed data in the left subfigure of Fig.~\ref{fig: RT}, logical constructs are not a “cost‑free” means of enhancing expressiveness; when introduced without sufficient constraints, they can substantially increase overall runtime. However, when appropriately integrated with basic syntactic constructs, the advantages of logical constructs within the specification generation framework can be fully exploited without incurring a significant increase in overall runtime. Under the modification paradigm, compared with CB, the corresponding runtime increases for the three configurations are 5.32\%, 14.48\%, and 1.85\%, respectively. This phenomenon arises for reasons similar to those observed for NVTC and is primarily attributable to the operational characteristics of the modification paradigm itself.

\begin{figure}[htbp]
  \centering
  
  \begin{minipage}{0.35\linewidth}
    \centering
    \includegraphics[width=\linewidth]{pic/proportion_of_program_with_optimal_conf_legend.pdf}
  \end{minipage}
  \vspace{-0.2em}

  \begin{minipage}{0.5\linewidth}
    \centering
    \includegraphics[width=\linewidth]{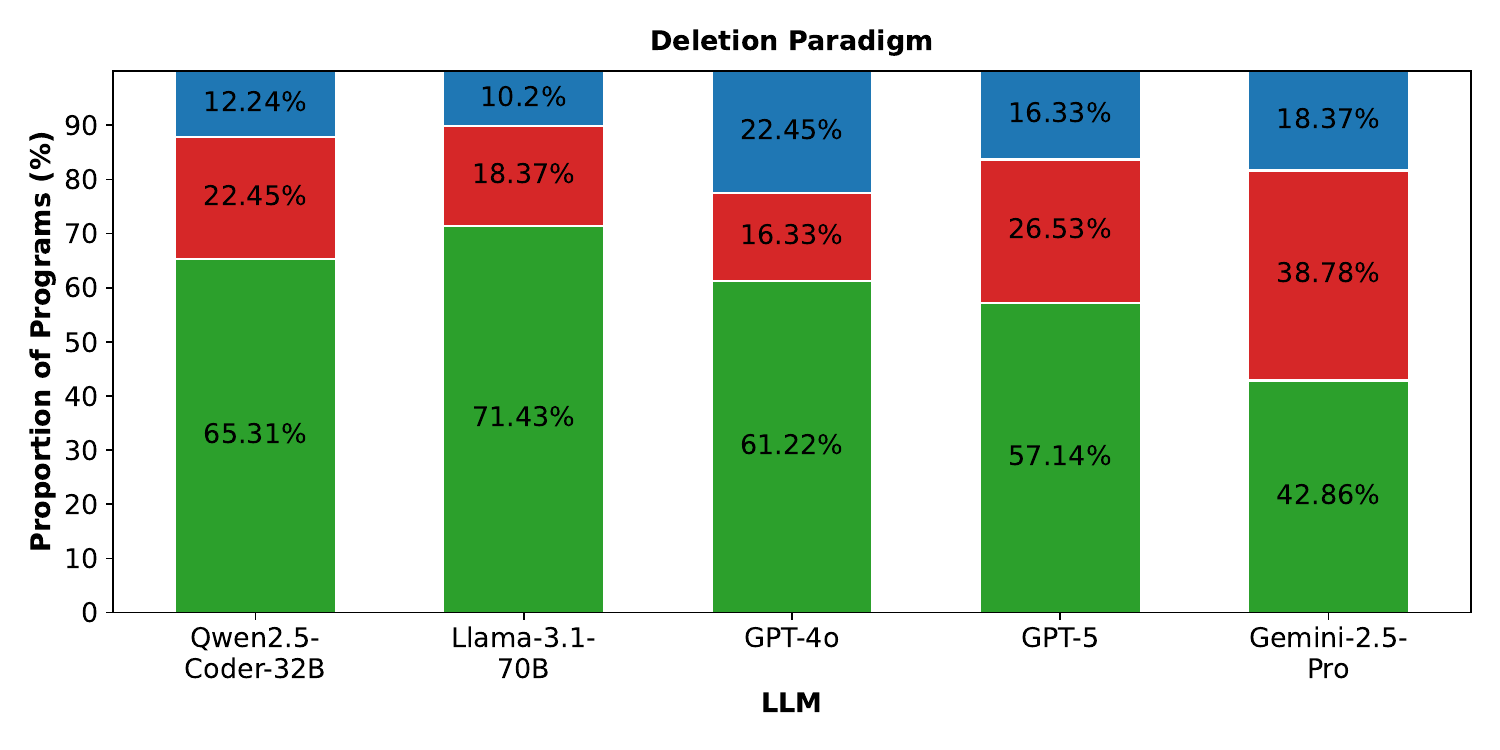}
  \end{minipage}%
  \begin{minipage}{0.5\linewidth}
    \centering
    \includegraphics[width=\linewidth]{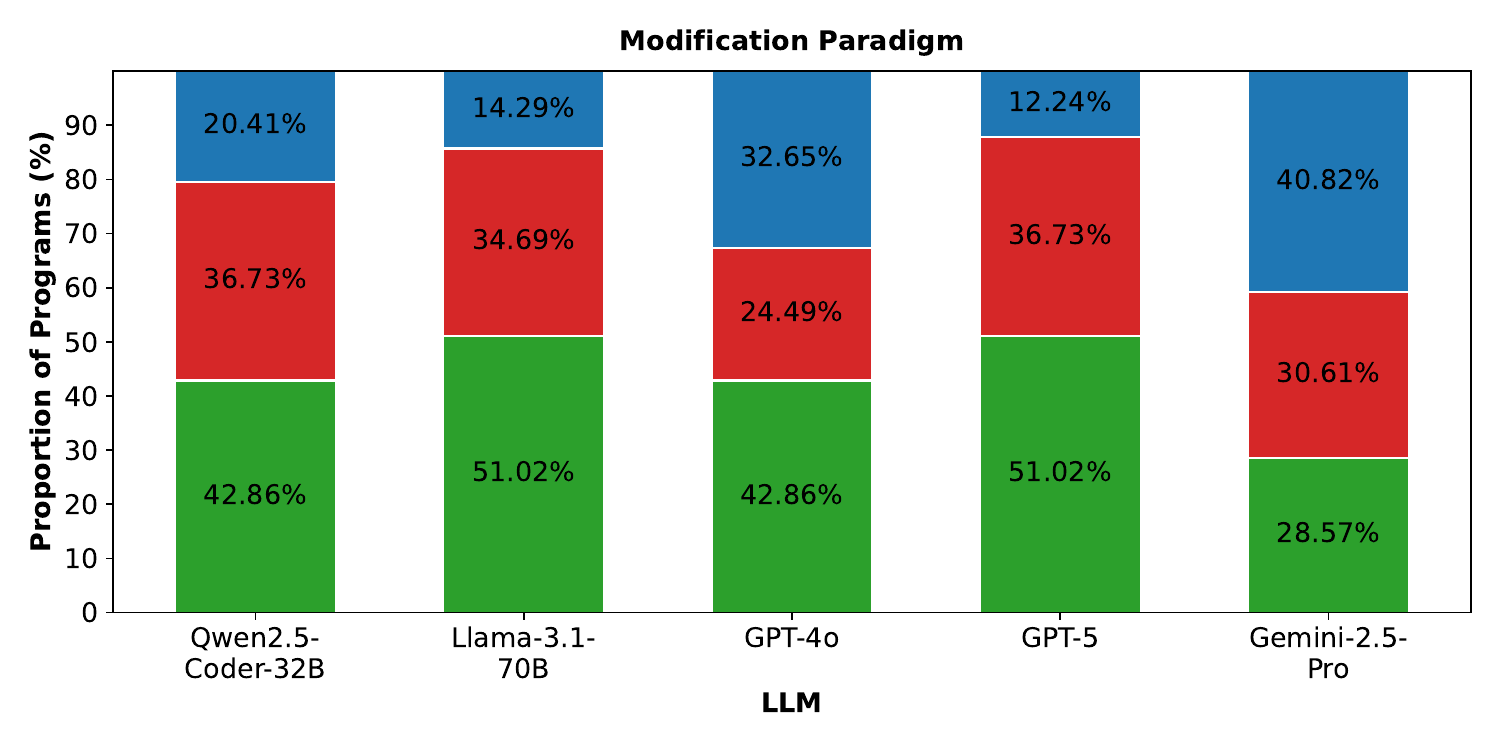}
  \end{minipage}%
  
  \caption{Proportion of Programs with Optimal Configuration (Achieving Minimum RT).}
  \label{fig: Proportion_RT)}
\end{figure}

Based on the distribution of configuration proportions shown in Fig.~\ref{fig: Proportion_RT)}, conclusions consistent with those drawn from Fig.~\ref{fig: Proportion_NVCT)} can be reached: each configuration corresponds to a distinct advantage regime with respect to the RT metric. Notably, as model capability increases, the dominance of the CB configuration gradually weakens, while configurations incorporating logical constructs become increasingly advantageous. This trend suggests that, with the support of higher‑capacity models, logical constructs are better able to fully realize their potential.

\begin{tcolorbox}[colframe=black, colback=gray!10, coltitle=black, boxrule=0.4mm]
    \textbf{Finding 4:} Although logical constructs increase verification complexity and runtime overhead to some extent, different syntactic constructs correspond to distinct efficiency advantage regimes. Consequently, when logical constructs are combined with basic syntactic constructs, verification capability can be enhanced without significantly increasing the exploration cost of the verification tool or the overall execution cost.
\end{tcolorbox}

\subsection{RQ3. Refinement Sensitivity}

As shown in Table~\ref{table: overall_performance}, with respect to the verification-capability–related metrics (NVP and NSVP), across the four configurations, the modification paradigm improves NVP by 17.61\%, 12.72\%, 16.17\%, and 6.49\%, respectively, relative to the deletion paradigm. The primary reason is that the deletion paradigm can only remove incorrect specifications; in doing so, it often discards useful but imprecise information as well, resulting in a weaker final specification. In contrast, the modification paradigm can leverage feedback to selectively supplement missing information, thereby substantially improving the verifiability of the resulting specification. Moreover, among the four configurations, CB exhibits the largest improvement, whereas CF shows the smallest. This is because CB permits only basic syntactic constructs, so the initially generated specifications are often insufficiently expressive. Because the deletion paradigm cannot recover missing content, the modification paradigm yields the greatest benefit under CB. Conversely, CF allows richer syntactic constructs from the outset, making it easier to generate relatively complete specifications even under the deletion paradigm; therefore, although the modification paradigm still yields gains, the marginal improvement is comparatively limited.

Consistent with the conclusions for NVP, NSVP under the modification paradigm is higher than that under the deletion paradigm across all four configurations. This indicates that, during iteration, the modification paradigm tends to steer specifications toward a state with a more coherent structure and more complete information, thereby reducing the impact of randomness in LLM generation on the final outcomes. However, the largest stability improvement in NSVP is observed for CA (32.09\%); the improvements for CV (20.00\%) and CB (19.73\%) are comparable; and CF (8.33\%) remains the smallest. Further analysis of the experimental samples shows that, even with the same LLM, specifications that include axiom constructs vary much more across sampling runs than those containing other logical constructs, and are therefore more prone to errors. Moreover, unlike the modification paradigm, the deletion paradigm cannot apply targeted fixes to axiom-related specification errors.

However, the verification capability gains brought by the modification paradigm come at the cost of higher verification overhead and runtime cost. Specifically, across all configurations, the modification paradigm exhibits higher NVTC, with the average increase ranging from 12.48\% to 31.83\%. Notably, CV shows the smallest increase, and its gap from the other configurations is relatively large. This is because, as analyzed in RQ2.3, configurations that enable verifiable logical constructs often require more failed attempts under the deletion paradigm, leading to higher NVTC than in other configurations. Under the modification paradigm, by contrast, a fixed upper bound on the number of iterations makes the NVTC values across the four configurations more similar, thereby yielding the smallest relative increase for CV. In addition, compared with the deletion paradigm, RT under the modification paradigm increases substantially across all configurations, with increases generally exceeding 100\%. This indicates that, although the modification paradigm offers clear advantages in verification capability, its end-to-end runtime also rises markedly. These increases mainly stem from the accumulated costs of multiple rounds of specification generation, verification feedback, and model inference.

\begin{tcolorbox}[colframe=black, colback=gray!10, coltitle=black, boxrule=0.4mm]
    \textbf{Finding 5:} There is no absolute superiority between the different refinement paradigms; rather, their applicability is contingent upon task objectives. In scenarios prioritizing verification efficiency and subject to time constraints, the deletion paradigm is more advantageous. Conversely, in scenarios pursuing higher verification coverage and specification quality, the modification paradigm offers stronger expressive and corrective capabilities.
\end{tcolorbox}

\section{Threat of Validity}
\textbf{Data Leakage Risk.} Although \textit{frama-c-problems} is an open-source dataset and likely included in the pre-training corpora of LLMs, this risk has a negligible impact on our experimental conclusions. First, the reference specifications in the original dataset are often incomplete and often fail verification. In our experiments, we modified the code context, forcing the LLMs to generate specifications adapted to the new context, thereby mitigating the risk of memorization in the baseline configuration (Config-Basic). Second, Config-Verifiable and Config-Axiom mandate the generation of specific logical constructs (such as lemmas and axioms) that are absent from the original dataset. Consequently, the results under these configurations reflect the synthesis capabilities of the LLMs rather than mere memorization of training data.

\textbf{Dataset Size Limitations.} Although the dataset comprises only 49 programs, it is widely recognized as a standard benchmark in the field of program verification. Despite its small size, the dataset features high algorithmic density, covering the most challenging verification patterns such as recursion, pointer manipulation, and array processing, thereby enabling an effective assessment of the logical completeness of the generated specifications. Given the current scarcity of large-scale, high-quality formal verification datasets, utilizing this benchmark ensures the comparability and reproducibility of our experiments. We leave the construction of a larger-scale verification dataset for future work.

\textbf{Cross-Language Generalizability.} Although this study is based on C/ACSL, we argue that our conclusions hold general applicability. First, the logical concepts underpinning each configuration (contract design, auxiliary lemmas, and axiomatization) represent universal paradigms in Hoare Logic and deductive verification, which are prevalent in specification languages such as JML~\cite{Burdy-01} and Dafny~\cite{Leino-01}. Our findings reflect the impact of varying levels of logical abstraction on verification, rather than being artifacts of specific language features. Second, we conducted multiple independent, repeated experiments across LLMs with diverse architectures. The consistency of the results further substantiates that the observed patterns stem from the general logical reasoning capabilities of LLMs, rather than from overfitting to specific language syntax.

\section{Related Work}
\subsection{Traditional Specification Generation}
Prior to the emergence of LLMs, specification generation primarily relied on Static Analysis (SA), Dynamic Inference (DI), and Data-driven Learning (DL). Through abstract interpretation~\cite{Cousot-01} or inductive reasoning, SA-based methods~\cite{Engler-01, Alur-01, Weimer-01, Ramanathan-01, Shoham-01, Nguyen-01, Albarghouthi-01} derive properties without executing the program. For instance, Flanagan et al.~\cite{Flanagan-01} proposed a technique for building an annotation assistant for modular static checker. However, these methods are often constrained by predefined abstract domains and templates, rendering them inadequate for capturing complex functional properties or developer intent. DI-based methods~\cite{Ernst-01, Ammons-01, Whaley-01, Yang-02, Gabel-01, Lee-01, Reger-01, Lemieux-01, Padhi-01} infer likely loop invariants by observing variable traces during program execution. For example, Ernst et al.~\cite{Ernst-01} developed Daikon, a dynamic invariant detection tool that supports program understanding and verification by automatically inferring invariants that likely hold based on runtime variable values. Although this approach is effective in detecting simple patterns, its effectiveness is strictly limited by the coverage of the test suite. To address the rigidity of the aforementioned rule-based systems, researchers have proposed DL-based methods~\cite{Krismayer-01, Mrowca-01, Si-01, Ryan-01, Yao-01, Liu-01, Yu-01, Si-02}. Frameworks such as Code2Inv~\cite{Si-02} and ICE-DT~\cite{Garg-01} leverage reinforcement learning or decision trees to synthesize invariants by formulating verification as a classification or search problem within a fixed grammar. However, these methods are typically confined to search spaces restricted by fixed grammars and are unable to express complex functional behaviors. Furthermore, they frequently generate obscure invariants that lack interpretability. In contrast, our work leverages the open-ended semantic understanding of LLMs, enabling the generation of expressive, human-readable specifications that align with developer intent.

\subsection{LLM-based Specification Generation}
The advent of LLMs has revolutionized the field of software engineering, prompting researchers to explore the integration of LLMs into program verification~\cite{Xie-01, Pei-01, Kamath-01, Wen-01, Ma-01, Chakraborty-01, Fan-01, Liu-02}. For instance, Xie et al.~\cite{Xie-01} conducted an initial evaluation of LLMs' capability to generate specifications from software annotations or documentation. However, our research focuses on generating specifications directly from program source code rather than software documentation. Pei et al.~\cite{Pei-01} investigated the potential of fine-tuning LLMs to generate program invariants. In contrast, our work emphasizes the utilization of existing (off-the-shelf) LLMs for specification generation, avoiding the need for model fine-tuning or retraining. Kamath et al.~\cite{Kamath-01} designed an early LLM-based automated specification generation framework, which employs prompt engineering to guide LLMs in generating whole-program specifications and utilizes verification tools to validate their correctness. Building upon this framework, Wen et al.~\cite{Wen-01} proposed AutoSpec, which leverages static analysis for hierarchical program decomposition and adopts a bottom-up approach to incrementally generate specifications. Ma et al.\cite{Ma-01} introduced an LLM-based automated specification generation technique for Java programs, utilizing a conversational approach to guide LLMs in producing appropriate specifications and filtering them through heuristic algorithms and mutation analysis. Furthermore, Wu et al.~\cite{Wu-01} proposed LEMUR, an automated theorem-proving framework that integrates LLMs with automated reasoners, while Liu et al.~\cite{Liu-02} presented ACInv, a tool specifically tailored for generating loop invariants in complex programs. However, none of the aforementioned studies incorporated the generation of auxiliary lemmas or axiomatization as a strategic component. Due to the absence of these logical abstractions, existing methods force underlying solvers to handle the entire proof complexity in a single step, which often leads to timeouts or verification failures in scenarios involving complex functional behaviors.

\section{Conclusion}
This paper presents the first systematic study exploring the feasibility and impact of introducing high-level logical constructs into LLM-based specification generation framework. Specifically, we propose a comprehensive evaluation framework comprising four configurations with varying levels of abstraction and conduct extensive empirical studies on multiple representative LLMs under two mainstream refinement paradigms. The experimental results highlight the critical role of the synergistic use of logical constructs and basic syntactic constructs in enhancing the verification capability and robustness of the framework.

\bibliographystyle{ACM-Reference-Format}
\bibliography{references}

\end{document}